\documentclass{aa}  

\usepackage{comment}
\usepackage{graphicx}
\usepackage{amsmath}
\usepackage{bm}
\graphicspath{ {figures/} }
\usepackage[dvipsnames]{xcolor}

\definecolor{red}{RGB}{219, 48, 122}
\usepackage{txfonts}
\usepackage[final]{hyperref}
\begin{document}

   \title{MaNGIA: 10,000 mock galaxies for stellar population analysis}
   
   \titlerunning{MaNGIA: 10,000 mock galaxies for stellar population analysis}

   \subtitle{}

   \author{Regina Sarmiento
                    \inst{1}$^,$
                    \inst{2}
          \and
           Marc Huertas-Company
                    \inst{1}$^,$
                    \inst{2}$^,$
                    \inst{3}
          \and
           Johan H. Knapen
                    \inst{1}$^,$
                    \inst{2}
          \and
           H\'ector Ibarra-Medel
                    \inst{4}$^,$
                    \inst{5}
          \and
           Annalisa Pillepich      
                    \inst{6}
          \and
           Sebasti\'an F. S\'anchez
                    \inst{7}
          \and
           Alina Boecker
                    \inst{1}
          }

   \institute{Instituto de Astrof\'isica de Canarias (IAC),              La Laguna, E-38205, Spain\\
            \email{regina.sarmiento@iac.es}
        \and
              {Departamento de Astrof\'isica, Universidad de La Laguna (ULL), E-38200, La Laguna, Spain}
        \and
              {Observatoire de Paris, LERMA, PSL University, 61 avenue de l'Observatoire, F-75014 Paris, France}
        \and 
              {Instituto de Astronom\'ia y Ciencias Planetarias, Universidad de Atacama, Copayapu 485, Copiap\'o, Chile}
        \and 
              {Escuela Superior de F\'{\i}sica y Matem\'aticas, Instituto Polit\'ecnico Nacional, U.P. Adolfo L\'opez Mateos, C.P. 07738, Ciudad de M\'exico, M\'exico}
        \and
              {Max-Planck-Institut für Astronomie, Königstuhl 17, D-69117 Heidelberg, Germany}
        \and 
              {Instituto de Astronom\'ia, Universidad Nacional Aut\'onoma de M\'exico, AP 70-264, CDMX 04510, Mexico}
}

   \date{Submitted 18 November 2022}

  \abstract
 {Modern astronomical observations give unprecedented access to the physical properties of nearby galaxies, including spatially resolved stellar populations. However, observations can only give a present-day view of the Universe, whereas cosmological simulations give access to the past record of the processes that galaxies have experienced in their evolution. To connect the events that happened in the past with galactic properties as seen today, simulations must be taken to a common ground before being compared to observations. Therefore, a dedicated effort is needed to forward-model simulations into the observational plane.
 }
 {We emulate data from the Mapping Nearby Galaxies at Apache Point Observatory (MaNGA) survey, which is the largest integral field spectroscopic galaxy survey to date with its $10,000$ nearby galaxies of all types. For this, we use the latest hydro-cosmological simulations IllustrisTNG to generate MaNGIA (Mapping Nearby Galaxies with IllustrisTNG Astrophysics), a mock MaNGA sample of similar size that emulates observations of galaxies for stellar population analysis.}
 {We choose TNG galaxies to match the MaNGA sample selection in terms of mass, size and redshift in order to limit the impact of selection effects. We produce MaNGA-like datacubes from all simulated galaxies, and process these with the stellar population analysis code pyPipe3D. This allows us to extract spatially resolved maps of star formation history, age, metallicity, mass and kinematics, following the exact same procedures used as part of the official MaNGA data release.}
 {This first paper presents the approach to generate the mock sample and provides an initial exploration of its properties. We show that the stellar populations and kinematics of the simulated MaNGIA galaxies are overall in good agreement with observations. Specific discrepancies, especially in the age and metallicity gradients in low- to intermediate-mass regimes and in massive galaxies' kinematics, require further investigation but are likely to uncover new physical understanding. We compare our results to other attempts to mock similar observations, all of smaller data sets.}
 {Our final dataset will be released with the publication, consisting of $\gtrsim10,000$ post-processed data-cubes analysed with pyPipe3D, along with the codes developed to create it. Future work will employ modern machine learning and other analysis techniques to connect observations of nearby galaxies to their cosmological evolutionary past.}

   \keywords{}

   \maketitle
%
\section{Introduction}

 Over the past years, we have witnessed an emergence of modern hydro-cosmological simulations  (e.g. \citealt{Eagle-1}, \citealt{Eagle-2}, \citealt{Illustris-1}, \citealt{Illustris-2}, \citealt{Illustris-3}, \citealt{Dylan-TNG100}, \citealt{Illustris-5}), which provide galaxies in a cosmological context and with unprecedented resolution, recovering structures in galaxies like spiral arms and bars (\citealt{TNG-FirstResults}). Characterizing how close are the properties of simulated and observed galaxies has become crucial to progress in our understanding of the physics of galaxy formation. 
  
  However, observations are affected by a number of instrumental and selection effects. A dedicated effort is necessary to emulate the observational effects in simulation outputs and produce statistical samples to be compared with observations. This apple-to-apple comparison is needed to validate whether the simulations reproduce the observed galaxies and their properties or not. Mock imaging has been extensively used to statistically compare simulated galaxies with observations delivered by large surveys as the Sloan Digital Sky Survey (SDSS, \citealt{York_2000}), Pan-STARRS (\citealt{Pan-STARRS}), identifying compatibilities between simulations and observations and leading to a new generation of improved simulations. Some examples include studies on galaxy morphology, colour distribution or mass to light relations (\citealt{Torrey-2015}, \citealt{Connor-2017-I}, \citealt{Connor-2017-II}, \citealt{Schulz-2020}, \citealt{Dylan-TNG100}, \citealt{vicente-statmorph}).
  
   In addition to comparing observations and simulations, realistic mock observations can also be used for simulation-based inference of physical processes which are accessible in the simulation, but not in observations at fixed redshift.
  
  In this work, we present MaNGIA (Mapping Nearby Galaxies with Illustris-TNG Astrophysics), a new forward-modelled sample from the Illustris-TNG simulation to mimic the properties of the MaNGA survey. MaNGA is currently the largest integral field spectroscopic survey with data for $10,000$ galaxies and enables a statistical analysis of spatially resolved physical properties of nearby galaxies. The survey's science goals include the study of galaxy assembly histories, galaxy quenching and its link to environment, and how galaxy morphology and components form \citep{MaNGA-Bundy}. 
  
  Previous works have addressed mock MaNGA datacubes for a reduced number of simulated galaxies, following different recipes. \cite{Connor-mock} emulated the instrumental effects of the survey with the \textsc{RealSimIFS} code and applied it to TNG50  simulations (\citealt{TNG-FirstResults}, \citealt{TNG-FirstResults-2}), producing mocks for $893$ simulated galaxies using input datacubes that comprise the stellar particles' line-of-sight velocity distributions. Other approaches produced full spectral datacubes prior to mimicking MaNGA's instrumental pipeline (\citealt{Hector-mock}, \citealt{Becky-mock}, \citealt{Lorenza-mock}). In these, the spectra were generated considering three components: stellar continuum, emission lines due to young stars and/or AGN feedback, and dust attenuation. \cite{Becky-mock} shows the impact of using the full mock datacubes to recover the kinematic maps.

  We present a set of $10,000$ post-processed datacubes that resemble the MaNGA survey, constituting the first complete sample and the largest one to date. We use TNG50 (\citealt{TNG-FirstResults}, \citealt{TNG-FirstResults-2}), the highest-resolution hydro-dynamical simulation of the IllustrisTNG family. This TNG simulation offers the best compromise to date between resolution and number of unique galaxies in a cosmological context. We mimic MaNGA's target selection and follow the procedures of \cite{Hector-mock} to generate mock datacubes from simulated galaxies.  These include producing the spectra associated to the particles in the simulation, recreating the fiber bundle observing scheme to mimic MaNGA's spectral and spatial resolution, and incorporating typical observational effects as atmospheric seeing and noise.  The data products are aimed to be comparable to the MaNGA MPL-11 release\footnote{\url{http://ifs.astroscu.unam.mx/MaNGA/Pipe3D_v3_1_1/index.html}} analyzed with the pyPipe3D code\footnote{\url{http://ifs.astroscu.unam.mx/pyPipe3D/}} (\citealt{pyPipe3D}, \citealt{pipe3d-10000}), which extracts the stellar populations and emission line properties from the integral field spectroscopy data.

 The paper proceeds as follows. Section\,\ref{Sec:samp_selec} details the sample selection strategy. The method used to generate mock MaNGA data-cubes is described in Section\,\ref{Sec:mocks}. In Section\,\ref{Sec:overview} we show an overview of the mock dataset. Data can be accessed by following the instructions in Section\,\ref{Sec:data_access}. Discussion and Conclusions are in Sections\,\ref{Sec:discussion} and \ref{Sec:conclusions}.

\section{Sample selection}\label{Sec:samp_selec}

\subsection{MaNGA}\label{Subsec:MaNGA-match}
 
 MaNGA (\citealt{MaNGA-Bundy}) is the integral field spectroscopic (IFS) survey of the Sloan Digital Sky Survey IV (SDSS IV; \citealt{SDSS-IV}). MaNGA targets were observed with the Sloan $2.5$\,m aperture telescope at APO \citep{MaNGA-Gunn}. To obtain the spatial-spectral cubes, MaNGA used integral field units (IFUs) formed by arrays of fibers distributed in a hexagonal pattern \citep{MaNGA-Drory}. The fibers are connected to two twin multi-object fiber spectrographs covering the $340-1,030$\,nm wavelength range with a spectral resolution $R\sim2,000$ \citep{MaNGA-Smee}. MaNGA data was released in $10$ public releases, completed in DR17 \citep{MaNGA-DR17}. 
 
 The MaNGA survey consists of over $10^4$ galaxies to have a statistical sample of galaxies that covers a variety of environments and star formation activity. To achieve this, the galaxies were selected to have a flat distribution in mass, as this parameter is an important driver of the galaxy population. However, to avoid uncertainties in the mass calculation, the selection was based on the i-band magnitude ($M_{\rm i}$) instead, commonly used as a proxy for stellar mass. A uniform distribution in $M_{\rm i}$ was achieved by selecting galaxies at different redshifts. 
 
 The complete sample is divided into three sub-samples: Primary, Secondary and Colour-Enhanced. The Primary sample represents $\sim50\%$ of the survey and optimizes a galaxy spatial coverage out to $1.5\,R_{\rm e}$, while the spatial coverage in the Secondary sample (designed to have half as many targets as the Primary sample) is out to $2.5\,R_{\rm e}$. The Colour-Enhanced sample was defined to balance the NUV-i colour distribution in the Primary sample by increasing the number of galaxies in the red, low-mass and in the blue, high-mass regimes and in the green valley. As the targets should have a similar spatial coverage ($1.5$ or $2.5\, R_{\rm e}$ of the galaxy), five bundles sizes were used to optimize the coverage. This introduces a constraint on the redshift and the size of the galaxies such that their apparent sizes match that of the bundles' field of view \citep{MaNGA-Bundy}. 
 
 To mimic the MaNGA sample we select simulated galaxies with similar redshifts, sizes and masses (see Sect.\,\ref{Sec:match_alg} for more details). In particular, we use the following parameters for the observed galaxies. The redshifts assumed are as listed in the NASA-Sloan Atlas (NSA) catalogue. As the MaNGA IFU allocation is based on the elliptical Petrosian $50\%$ light radius in the SDSS $r$-band, we use this parameter as a proxy for size. Although the original sample selection replaces stellar mass by the $i$-band magnitude, deriving magnitudes from simulated galaxies involves a mocking process itself which is computationally very expensive and not justified at the target selection stage of this work. Therefore we use the stellar mass $M_{\rm obs}$ within an aperture of two Petrosian radii. As the redshifts, the effective radii and stellar masses are extracted from the NSA catalogue.

\subsection{TNG50}\label{Subsec:TNG-match}

 TNG50 uses a finite periodic co-moving volume of $50\,{\rm Mpc}^3$ and assumes a flat ${\rm \Lambda CDM}$ cosmology with $H_0=67.74\,\,{\rm km\,s}^{-1}\,{\rm Mpc}^{-1}$ \citep{cosmo-planck}, which is the one used throughout this work. As the cosmological box evolves through time, $100$ snapshots are saved with a typical separation of $\Delta z=0.01$ across MaNGA's redshift range, which is equivalent to a mean time-span of $0.16$\,Gyr between snapshots. We consider simulated galaxies from snapshots 87 to 98 ($0.012\leq z \leq 0.15$).

 While MaNGA targets have been selected based on properties derived from the SDSS survey, photometric measurements are not directly derivable from the simulations. Therefore, the challenge of matching observed galaxies with simulated ones relies on their properties being comparable. While redshifts are defined by the TNG snapshots, radius and mass require a more detailed estimation. We use the preexisting catalogues from \cite{vicente-statmorph} which collect photometric parameters of TNG50 galaxy mocks produced using the \textsc{SKIRT} radiative transfer code (\citealt{Skirt-1}, \citealt{Skirt-2}). As these catalogues are only available for a limited number of TNG snapshots, we aim to relate the radii directly calculated from the simulations (i.e. the stellar half-mass radii) with the photometric radii derived from the mocks to later extend this relation to all the snapshots with $z\leq0.15$. In particular, we use the catalogue corresponding to snapshot $95$ ($z=0.048$) as it is the one with the nearest redshift to MaNGA's average redshift that is available. Galaxies in this snapshot are divided in five stellar mass bins to later perform a linear fit per bin between the simulation and observation-like sizes. An analogue procedure is followed to estimate the circularized Petrosian radius, which is then used to calculate the stellar mass of the galaxy enclosed within two Petrosian radii. In Appendix\,\ref{Ap:r-fit} we describe in detail the fit performed to estimate the elliptical Petrosian half-radii and the stellar masses for the simulated galaxies. 
 
 As other parameters such as colour, star formation rate (SFR) and environment are not matched, the TNG-MaNGA counterparts will not necessarily share the exact same properties. However, both the MaNGA sample and our TNG sample should have a similar variety of these properties, as the target selection is essentially the same. We avoid matching the samples by colour as the magnitudes derived from the simulation are not necessarily comparable to the observed ones. Although the Colour-Enhanced sample is colour-defined, this sample represents only $\lesssim15\%$ of the total. Nevertheless, we identify TNG50 analogues for these galaxies, but only considering the parameters described before. This means that this sub-sample of MaNGA may not be represented as well as the others in our sample. However, the simulated galaxies are \emph{observed} with the largest bundle size to enable a posterior rearrangement of the samples.

\subsection{Matching algorithm}\label{Sec:match_alg}

  Our simulated sample is built by finding a TNG50 analogue to every MaNGA galaxy, matching redshift, stellar mass and $R_{\rm e}$ with a similar approach to \cite{mangalike-sample}. We use the DRPall catalogue v-3.1.1 corresponding to SDSS Data Release $17$. In this catalogue, $10,233$ observed galaxies have a well-defined mass, $R\rm{_e}$ and redshift, of which $10,094$ are unique. We consider all the galaxies with $M_*>10^{8.5} M_{\odot}$ in this sub-sample for the matching of the sample ($10,070$ in total). 

  As the number of simulated galaxies is limited, we make two assumptions that allow us to complete the sample. First, galaxies have modified their properties enough across redshift such that they can be considered as different galaxies if they are in different snapshots. Second, mocks produced by observing a simulated galaxy from different angles are sufficiently different to be considered as independent galaxies. However, the latter assumption fails for galaxies with a near to spherical symmetry or a large number of viewing angles. To reduce the impact of this assumption, the selection strategy should not only seek to minimize the difference in mass, size and redshift between the observed and simulated counterparts, but it should also maximize the assignment of different TNG50 galaxies. Therefore, we first perform a match considering only unique galaxies and later allow repetition until the sample is complete. The repeated galaxies are then \emph{observed} from different angles.

  For the match, each MaNGA galaxy is assigned to the TNG50 snapshot that has the nearest redshift to that of the observed galaxy. The TNG50 galaxy counterpart for the observed one must be in this snapshot. This limits the TNG50 snapshots used to $87-98$, with redshifts from $0.012$ to $0.15$. We then define a distance $d$ in the mass-$R_{\rm e}$ space as 

\begin{equation}
\begin{split}
    d^{2} = \left[\log(M_{\rm obs}[M_{\odot}])-\log(M_{\rm{TNG}}[M_{\odot}])\right]^2 + \\ 
    \left[\log(R_{\rm e,\,\,obs}[\arcsec])-\log(R_{\rm e,\,\,\rm{TNG}}[\arcsec])\right]^2\label{Eq:dist}
\end{split}
\end{equation}

\begin{figure*}
    \centering
    \includegraphics[width=\hsize]{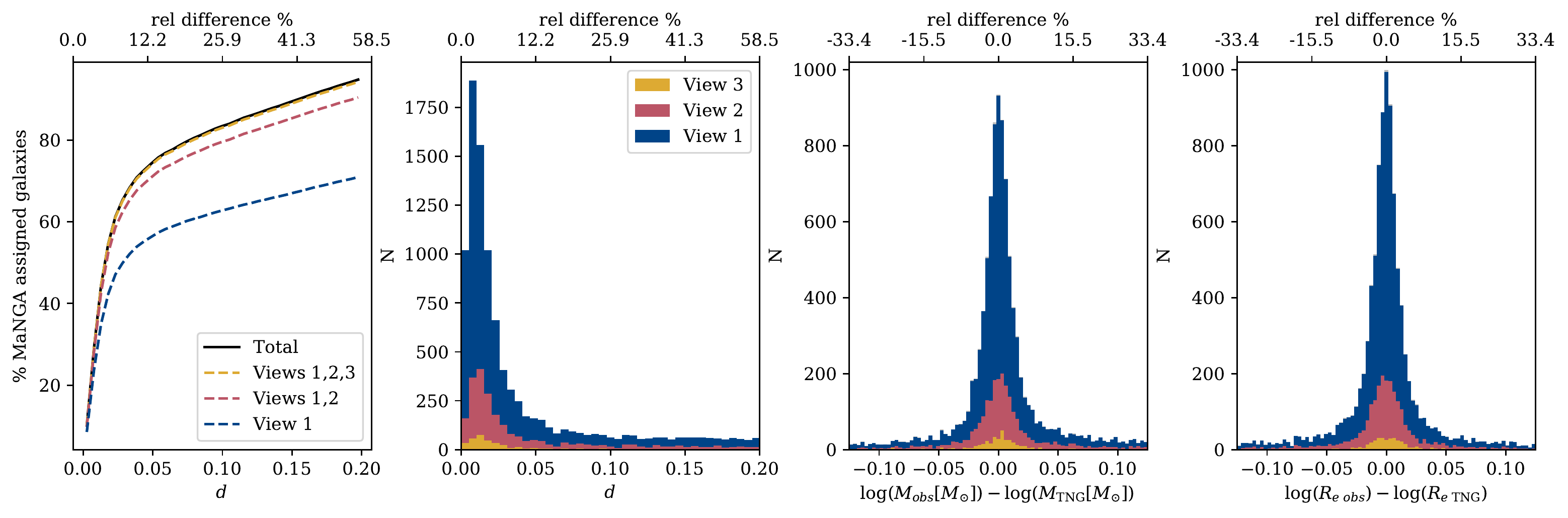}
    \caption{Distances in the mass-radius plane between MaNGA galaxies and their assigned TNG galaxy pair. From left to right: cumulative percentage of galaxies at a distance given by Eq.\,\eqref{Eq:dist}. Histogram of distances with galaxies paired with three repetitions. Histograms of the differences in the mass and radius components in $d$. The majority of MaNGA galaxies are allocated a TNG pair that is within a relative difference of $25\%$ in the mass-radius plane.}
    \label{fig:samp_dist}
\end{figure*}

 where masses and radii from observed and simulated galaxies are defined as in Sections\,\ref{Subsec:MaNGA-match} and \ref{Subsec:TNG-match} and Appendix\ref{Ap:r-fit}. Masses $M$ and radii $R_{\rm e}$ are in log-scale such that the difference between observed and simulated parameters is relative (i.e. setting an upper limit to $d<0.1$ implies that mass and radius are constrained to have relative differences smaller than $25\%$). 

 Each MaNGA galaxy gets paired with the nearest TNG50 galaxy in the $M-R_{\rm e}$ plane. As one TNG50 galaxy might be the nearest neighbour of multiple MaNGA galaxies in the mass-size plane, we define the following iterative algorithm to choose priority pairs. If a TNG50 galaxy gets selected more than once (for example $n$\,times) then the nearest MaNGA galaxy is assigned to this TNG50 galaxy, which will no longer be available to be selected in later iterations. The $n-1$ remaining galaxies will be sent back to the pool of unassigned MaNGA galaxies. This step is repeated until there are no more TNG50 galaxies available within the imposed range. We find that $7,522$ MaNGA galaxies get assigned a unique TNG50 match if $d$ is limited to $d<0.25$, which is of the order of the expected errors in the mass and radius estimations from observations.

 Because after this step a significant fraction of the MaNGA sample remains unassigned, we allow TNG50 galaxies to be repeated. This is done by repeating the previous step until all MaNGA galaxies are allocated, with the consideration that the simulated galaxies selected in each new iteration of this step are repetitions of previously selected galaxies. This results in $99.2\%$ of MaNGA galaxies assigned with up to three repetitions per TNG50 galaxy. Only $45$ TNG50 galaxies need to be repeated more than three times. Only ten MaNGA galaxies remain unassigned and four are assigned to TNG50 galaxies that have been chosen already six times. These $14$ galaxies are outliers in the $M_*-R_{\rm e}$ plane and therefore are left unassigned in our sample.
 
 The repeated TNG50 galaxies are \emph{observed} from different angles. The galaxies that appear in the sample less than four times are observed in the direction of the $x$, $y$ and $z$ positive axes, while those that appear four or to six times are observed from up to six isotropically distributed directions.
 
 The resulting distribution of distances $d$ in the $M-R_{\rm e}$ plane is shown in the first and second panel of Fig.\,\ref{fig:samp_dist}. We highlight that the mass and radius terms have similar distributions, indicating that neither term systematically dominates the final value of $d$ (Fig.\,\ref{fig:samp_dist}, right-hand side panels). Furthermore, both distributions are roughly symmetric and centred around zero, suggesting that the simulated and observed parameters populate the same regions of the M-R plane and do not introduce a further bias (that could have happened since one sample is mass-limited, while the other are a series of volume-limited samples, one per snapshot).

\begin{figure*}
    \centering
    \includegraphics[width=\hsize]{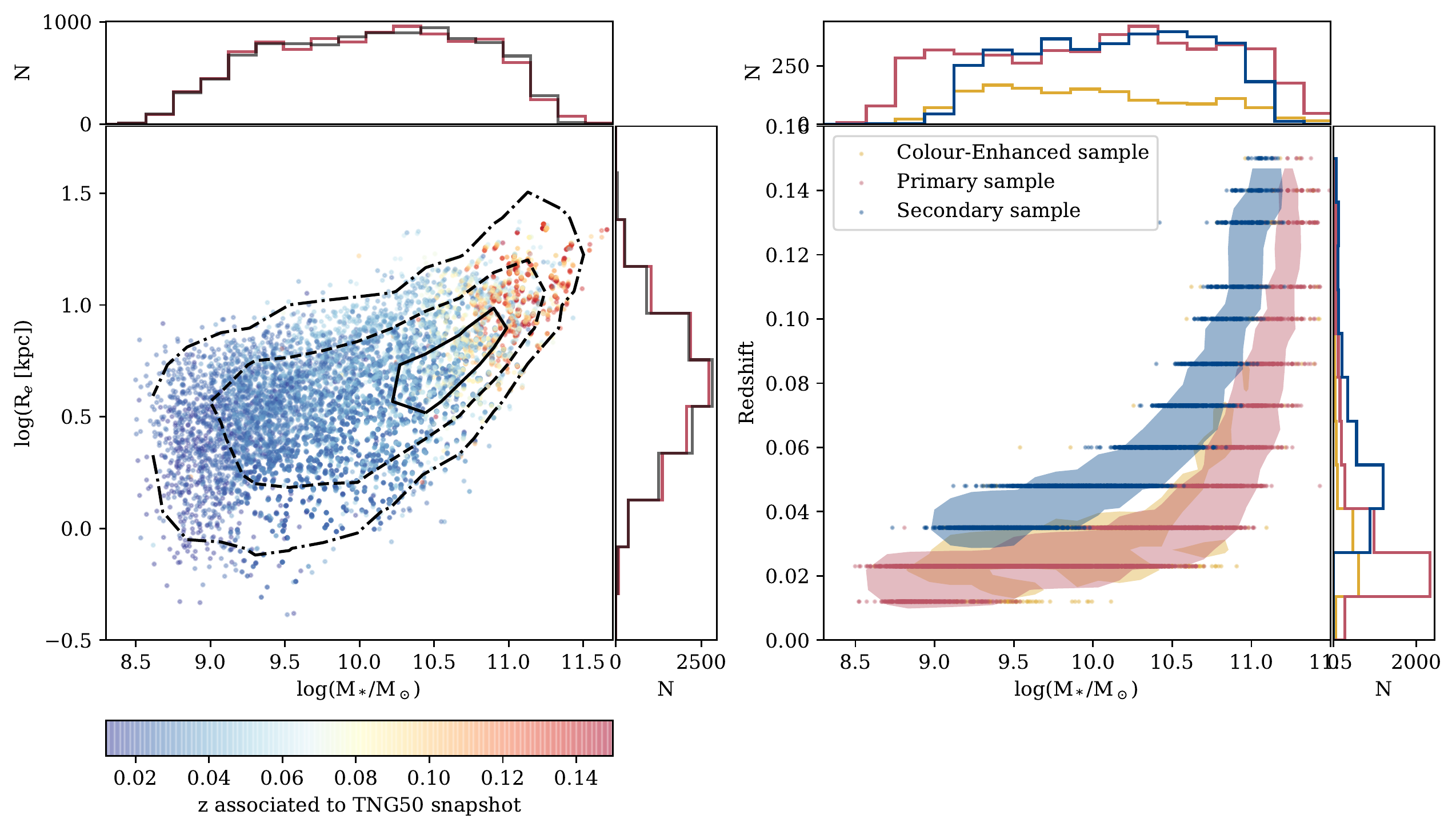}
    \caption{\emph{Left panel:} The distribution of mass and effective radius ($M_*-R_{\rm e}$ plane) of the TNG galaxies in the matched sample, colour-coded by redshift. The black contours indicate the density of galaxies in the MaNGA $M_*-R_{\rm e}$ plane.
    \emph{Right panel:} Mass vs. redshift of the TNG galaxies in the matched sample. Colour-coding indicates the MaNGA sub-samples: galaxies in the \emph{Primary sample} in red, \emph{Secondary sample} in blue and \emph{Color-Enhanced sample} in yellow. The contours show the distribution of the MaNGA galaxies in the plane.}
    \label{fig:match-M-z-R}
\end{figure*}

 In addition to analysing the distance distribution of the pairs, we compare the distribution in mass, size and redshift of MaNGA galaxies to that of the matched sample. Because every TNG50 snapshot is associated with a redshift, the TNG50 galaxies in our sample have a discrete distribution of redshifts that reflects the twelve snapshots used for the match (see the right panel in Fig.\,\ref{fig:match-M-z-R}). Although the densest regions in the $M-R_{\rm e}$ plane are comparable, some outliers cannot be matched in the TNG50 sample, for instance, the large galaxies in the massive end, as seen in Fig.\,\ref{fig:match-M-z-R}, left panel.  In these regions of the plane repetitions are more frequent because TNG50 is a volume-limited sample. As the number of galaxies per mass decreases toward high masses, fewer massive galaxies are available. Another region where our sample appears to be less dense is at small $R_{\rm e}$, however, all MaNGA galaxies in this region are paired. The effect can then either be due to repetitions or to galaxies that are selected in multiple snapshots but do not vary their properties across time.  The mass distribution is properly reproduced.

\section{Construction of a mock dataset}\label{Sec:mocks}

\begin{figure*}
    \centering
    \includegraphics[width=\hsize]{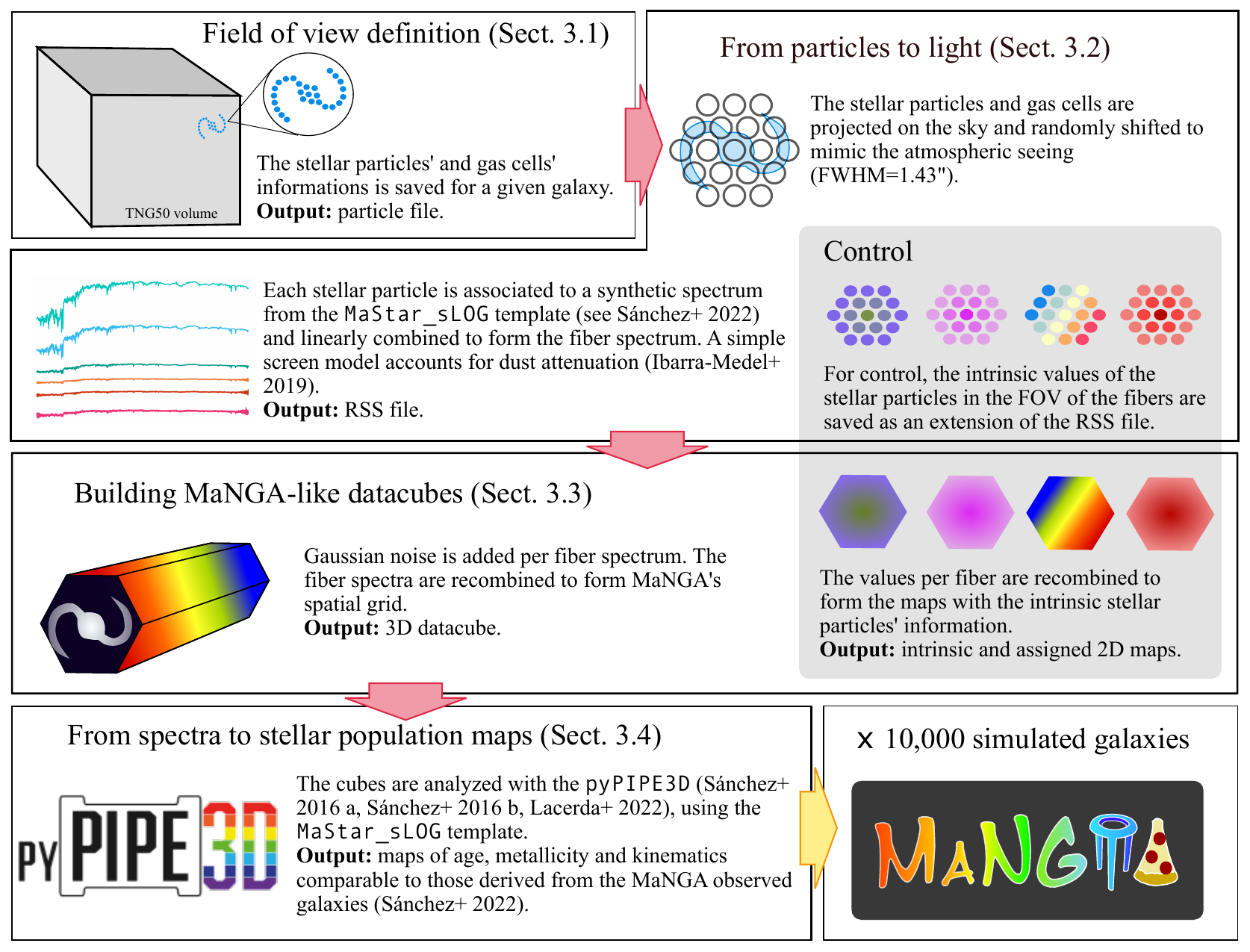}
    \caption{Scheme followed to build a MaNGA-like datacube from a TNG50 simulated galaxy. Each step is described in Sect.\,\ref{Sec:mocks}}
    \label{fig:method-scheme}
\end{figure*}

Once the subset of TNG50 galaxies has been selected (Sect.\,\ref{Sec:samp_selec}), a mock MaNGA-like datacube is constructed for each of the simulated galaxies in this sample. We describe the steps for generating mock datacubes below. We first define the field of view (FOV) of the observations. This limits the particles and cells from the simulation that will contribute to form the spectra in the final datacube. Latter, instrumental effects are included to emulate MaNGA observations. The procedure is based on \cite{Hector-mock} and further details can be found there. A scheme of the steps followed is shown in Fig.\,\ref{fig:method-scheme}.

\subsection{Field of view definition}\label{Subsec:mocks-FOV}

\begin{figure}
    \centering
    \includegraphics[width=0.9\hsize]{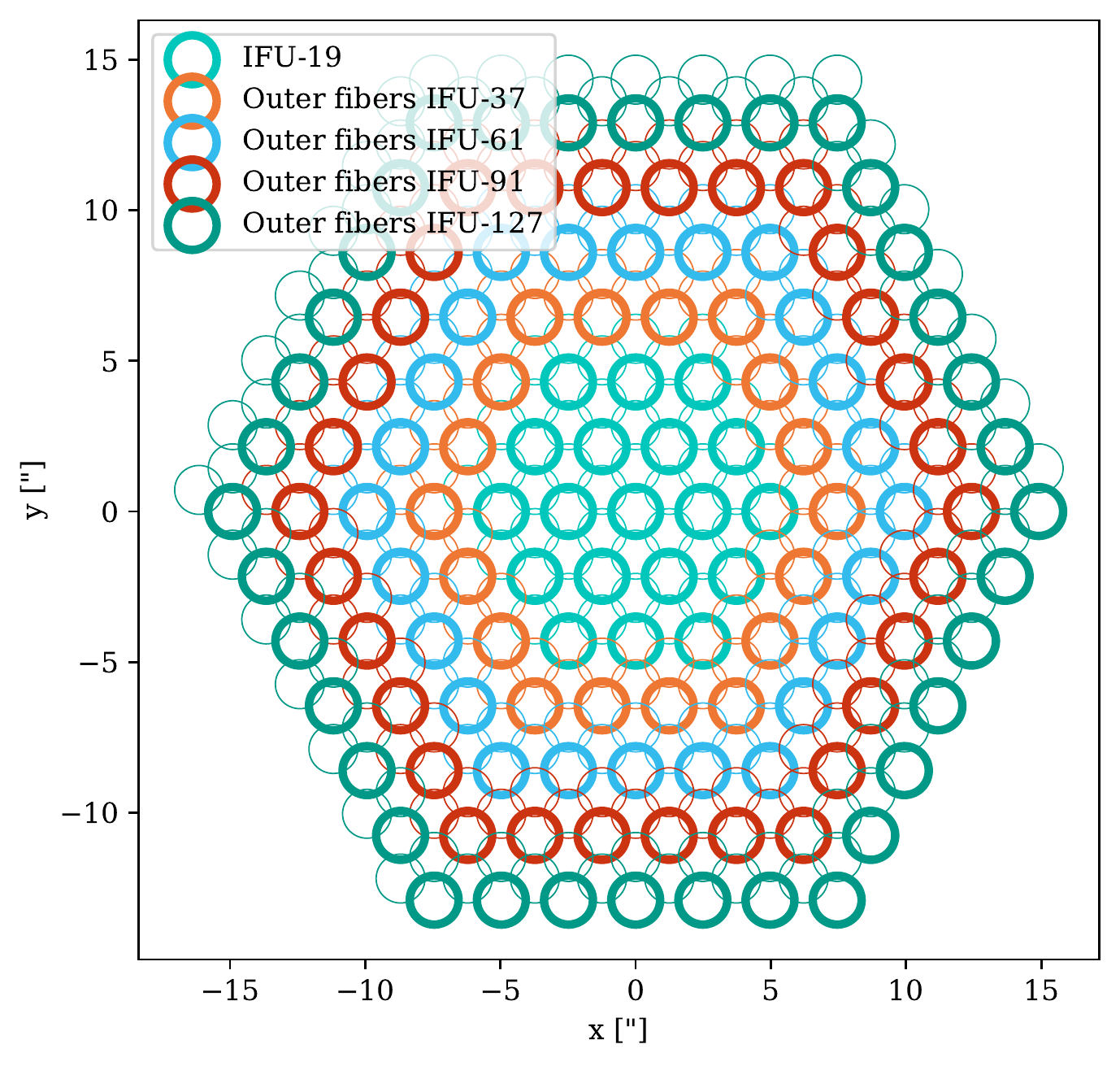}
    \caption{Fiber arrays used by MaNGA. Outer fibers of each MaNGA IFU bundle are shown in a different colour and a thick line, while circles with a thinner line width show the other two dithers for the same IFU. Large bundles have the same fiber array than smaller bundles, with additional rings of fibers in the outer parts. The inner-most fibers (IFU-19, light green) are shared by all the bundle sizes.}
    \label{fig:ifu-rings}
\end{figure}

The MaNGA survey uses IFUs formed by arrays of fibers distributed in a hexagonal pattern (Fig.\,\ref{fig:ifu-rings}). The bundles have five different sizes with diameters $12.5\arcsec$, $17.5\arcsec$, $22.5\arcsec$, $27.5\arcsec$ and $32.5\arcsec$. These correspond to IFUs of $19$, $37$, $61$, $91$ and $127$ fibers, respectively. To form a MaNGA datacube, the target is observed with one of these IFUs in three dithering positions resulting in a series of spectra associated with the fiber position in the FOV \citep{MaNGA-Drory}. Later, the fiber-spectra are recombined and sampled onto a grid of $0.5\times0.5\,\rm{arcsec}^2$. 

The survey is designed such that every galaxy has a similar spatial coverage. Each galaxy is allocated to the bundle size that optimizes either $1.5$ or $2.5\,R_{\rm e}$ coverage if they have been chosen for the Primary or the Secondary sample, respectively \citep{MaNGA-Wake}. Because the galaxies in our sample are matched in size and redshift, the proportion of galaxies assigned to each bundle is kept. However, to enable different sample selections, all mocks are produced with the largest bundle as smaller bundles can be obtained by removing the outer fibers.

The line of sight to the galaxy is defined by an unitary vector and the centre of the galaxy as listed in the TNG catalogue ("SubhaloPos"). For those galaxies that require up to three repetitions, we use the unitary vectors parallel to $x$, $y$ and $z$ axes: $\hat{\boldsymbol{\imath}}$, $\hat{\boldsymbol{\jmath}}$ and $\hat{\bm{k}}$ for views $1$, $2$ and $3$, respectively. When the TNG galaxy requires more than three repetitions, we define six isotropically distributed views as in Appendix\,\ref{Ap:iso-views}. The orientation of the galaxy with respect to the observer is random as it is given by the position of the particles in the simulation cube (see more details in Appendix\,\ref{Ap:iso-views}). The galaxy is \emph{observed} from the distance defined by the redshift of the snapshot the galaxy was taken from.

We consider all the stellar particles and gas cells from the simulation collected by the Friends-Of-Friends (FOF) algorithm. The FOF algorithm groups particles together if they are within a given linking length of each other, or of another particle already linked to the group. In the TNG simulation, these groups trace the dark matter haloes. They may include more than one galaxy and, in such a case, they are understood as a central galaxy and its satellites \citep{halo-FoF}. This means that not only the particles/cells to the target galaxy are considered, but also those of the neighbouring galaxies which reproduces a more realistic environment. 
 
As the spatial axes of the cube are formed by the combination of multiple fiber observations, each fiber will have its own FOV. Atmospheric seeing is emulated by applying a random shift to the simulation's particle positions. The shift follows a multivariate Gaussian distribution with ${\rm FWHM}=1.43\arcsec$. This shift is updated for each fiber and determines which particles fall in the field of view of the fiber.

\subsection{From particles to light}\label{Subsec:mocks-spec}

\begin{figure*}
    \centering
    \includegraphics[width=\hsize]{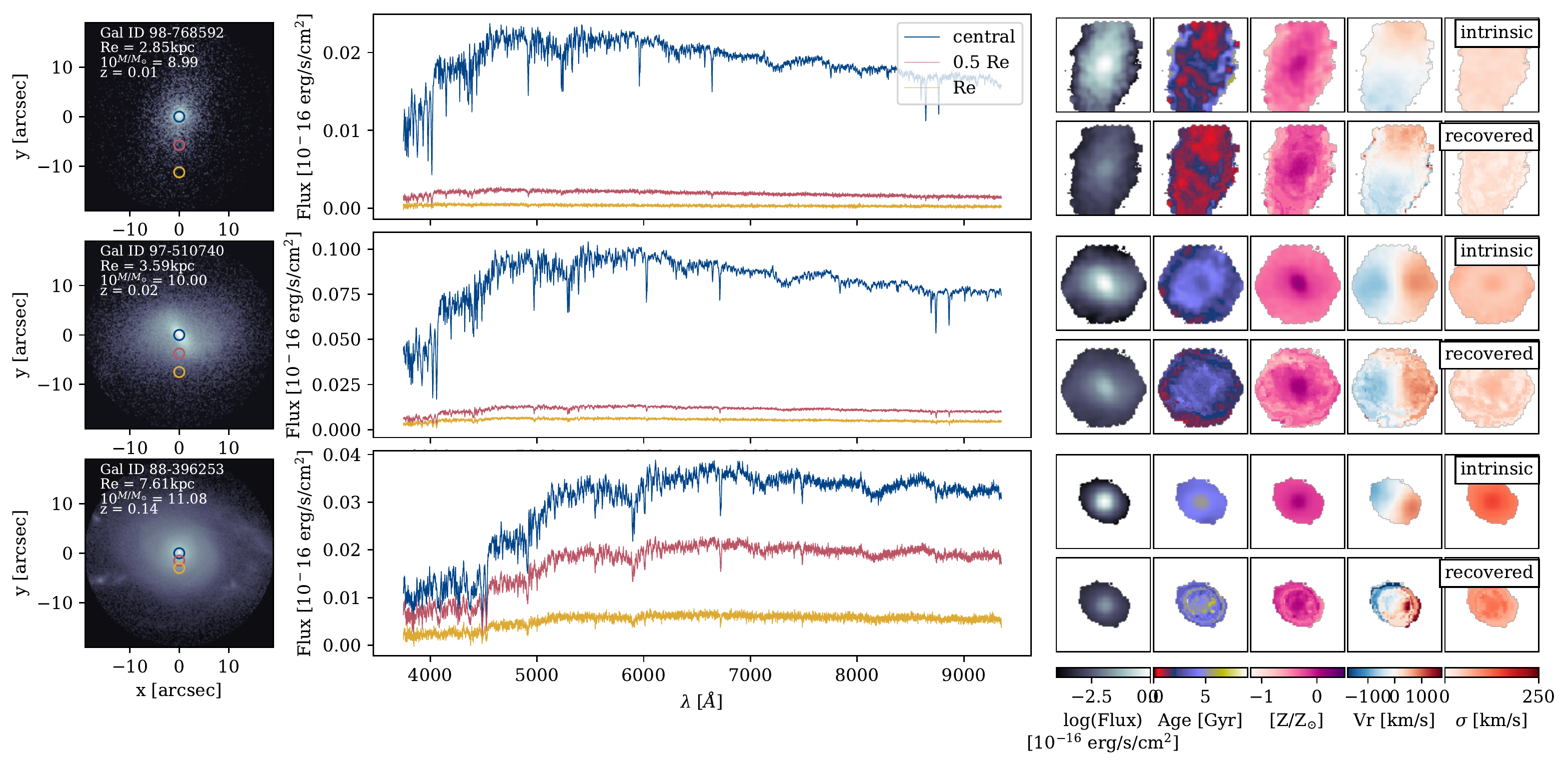}
    \caption{Three simulated galaxies with stellar masses $10^{9}M_{\odot}, 10^{10}M_{\odot}$ and $10^{11}M_{\odot}$ show the mocking process. Panels on the left-hand side show the stellar mass density of the galaxies, while the central panels show their spectra at different radii (centre in blue, $0.5R_{\rm e}$ in red and $1R_{\rm e}$, in yellow). Squared panels on the right-hand side show the stellar properties derived directly from the particles (top row for each galaxy) and recovered from the spectral cubes with \textsc{pyPipe3D} (bottom row for each galaxy). From left to right, these panels show the $V$-band reconstructed image (mass density for the particle maps), luminosity-weighted age, luminosity-weighted metallicity, LOS velocity and velocity dispersion. As galaxy mass increases, the spectra are redder. Stellar maps are, in general, properly recovered.}
    \label{fig:spec_at_R}
\end{figure*}

The mock spectra are generated by including the stellar emission associated with the simulated stellar particles and the absorption produced by the presence of dust in the line of sight. Gas emission lines produced in H{\sc ii} regions are simulated by associating star-forming gas cells to \textsc{Cloudy} templates (see \citealt{Hector-mock}). However, because the interstellar medium model behind IllustrisTNG (\citealt{Springel-Hernquist-2003}) does not allow us to rely on the temperature and other thermo-dynamical properties of the star-forming gas, we remove the emission line contribution from the spectra for our analysis. As \textsc{pyPipe3D} performs two spectral fits to extract emission lines before fitting the simple stellar population (SSP) template, the analysis of our spectra should be comparable to that performed on observations. Nonetheless, a non-perfect subtraction of the emission lines or the superposition of emission with absorption lines introduces an additional noise to the spectrum that could affect the template fitting. We evaluate this effect in Appendix\,\ref{Ap:emlines}. 

The SSP template was chosen with the following considerations. On the one hand, the SSP template used to create the mock cubes should be the same as the one used in the stellar population analysis (see Sect.\,\ref{Sec:SSP-ana}). This is to avoid introducing uncertainties due to the assumptions behind the stellar populations template. On the other hand, to be consistent with the observations, the stellar population analysis should be carried out with the same template used to process MaNGA observations. Therefore we adopt the SSP template used in \cite{pipe3d-10000}, {\tt MaStar\_sLOG}.

The {\tt MaStar\_sLOG} template consists of a set of synthetic spectra generated with the \textsc{Galaxev} code (\citealt{bruzual-charlot}, \citealt{Yan_2010}) based on MaNGA's stellar library, MaStar \citep{MaNGA-Yan}. This library has spectra of a variety of single stars, in order to cover a wide range of effective temperature, surface gravity, metallicity and alpha-elements-to-iron ratio. \textsc{Galaxev} combines these spectra assuming a \cite{Salpeter} initial mass function (IMF) and \textsc{Parsec} isochrones \citep{PARSEC} to produce synthetic stellar populations spectra. In particular, we use a subset of $273$ synthetic spectra $L_{\rm SSP}(\lambda, Z, t)$ comprising $39$ ages between $0.0023 \le t \le$ $13.5$\,Gyrs with the \texttt{sLOG} sampling, seven metallicities  $0.0001 \le Z \le 0.4$ \footnote{A solar metallicity of $Z_{\odot}=0.01698$ was adopted throughout the paper to covert $Z$ to $Z/Z_{\odot}$ units and square brackets are used to indicate log-scale, $\log(Z/Z_{\odot}) = [Z/Z_{\odot}]$.} (see \citealt{Artemi-SSP} for more details of the template age-metallicity sampling) and a linearly sampled wavelength range of $2,000-10,000\,\AA$ at rest. Because the MaStar library was obtained using the same instrument as the one used for the MaNGA survey, the spectra have the same spectral resolution.

Because the data cubes are obtained imitating the IFU bundles, the spectra are constructed per fiber with the information of all the particles within the FOV of the fiber. From now on, we will refer to these spectra as \emph{fiber spectra} or $F_i(\lambda)$, the $i-$th fiber spectrum in a bundle.

The $n_{\rm s}$ stellar particles in the field of view of each fiber in the bundle contribute to form the spectrum. The $j-$th stellar particle is assigned the SSP with the nearest age and metallicity of the template. The spectrum is then mass-weighted and normalized by the mass-luminosity factor $ML_{\rm {SSP}}$ associated to the corresponding SSP to obtain the flux. Latter, the spectrum corresponding to the $j-$th stellar particle as a function of wavelength is given by

\begin{equation}
\begin{split}
    F_j^{\rm \,stellar}(\lambda, Z_j, t_j, m_j, r_j) &= \\ 
    L_{\rm {SSP}}\,(\lambda,\, Z_j,\, t_j)& \times \frac{m_j}{ML_{\rm {SSP}}(Z_j,\, t_j)} \times \frac{L_{\odot}}{4\pi r_j^2}\,\,\,,
\end{split}
\end{equation}

where $r_j$, $Z_j$, $t_j$ and $m_j$ are the distance, metallicity, age and mass of the $j-$th stellar particle.

The presence of dust between the emitting source and the observer produces absorption in the optical range. As the simulation does not include dust particles, the dust content of the galaxies is estimated as a gas fraction following the recipe of \cite{Remy-dust}, which is metallicity-dependent. The effect of dust is introduced in the mocks considering a simple screen model that attenuates the spectrum associated to every single stellar particle using the relation $A_{\lambda}/A_V$ from the \cite{Cardelli} extinction law with an extinction factor $R_V=3.1$. For this purpose, the extinction at Johnson's $V$ photometric band $A_{V, j}$ is calculated for the $j-$th stellar particle as the contribution of all the gas cells that are within the fiber's FOV and placed between the observer and the stellar particle (see \citealt{Hector-mock}). Then, the attenuated spectrum of the $j-$th stellar particle is 

\begin{equation}
  F_j^{\rm \,stellar\,\,obs}(\lambda) = F_j^{\rm \,stellar}\times 10^{-0.4 \frac{A_{\lambda}}{A_V} A_{V,j}}.
\end{equation}

The spectrum is then Doppler-shifted in wavelength according to the particle's velocity in the line of sight $v_j$ and the redshift $z$ at which the galaxy is placed. The spectrum of the $i-$th fiber is calculated as the sum of all the stellar contributions as

\begin{equation}
    F_i(\lambda) = \sum_{j}^{n_s} F_j^{\rm \,stellar\,\,obs}\left(\lambda+ \lambda\times\left[\frac{v_j}{c}+z\right]\right) \,\,,
\end{equation}

where $c$ is the speed of light. We highlight that because the emission associated to each stellar particle is individually attenuated, the dust extinction in different regions of a galaxy can be different. 

Even though the dust model used in this work is rather simple, it is not less complex than the reverse analysis that \textsc{pyPipe3D} performs. Other more complex approaches include radiative transfer (\textsc{Skirt}, \textsc{Sunrise}), where the gas cells are treated as a Voronoi grid which is compatible with how the simulation is computed. These approaches are considerably more expensive computationally than ours.

At this point the fiber spectra are saved in the row-stacked spectra (RSS) format. For control, the intrinsic properties of the stellar particles in the FOV of the fibers are saved as an extension of the RSS file. The luminosity-weighted (LW) values of age and metallicity, are computed using the {\tt MaStar\_sLOG} and weighting by the flux at $5,000\,\AA$. To be consistent with the stellar analysis applied to the datacubes, these are added in log-scale for the $i-$th fiber as

\begin{equation}
  \begin{split}
    \log{\rm (Age)}_i &= \frac{\sum_j^{n_s} w_j \log{\rm (t_j)}}{\sum_j^{n_s} w_j}\,\,,\\
    \log{\rm (Z)}_i &= \frac{\sum_j^{n_s} w_j \log{\rm (Z_j)}}{\sum_j^{n_s} w_j}\,\,,\, {\rm and}\\
    \\
    w_{j} &= \frac{m_j}{ML_{\rm {SSP}}(Z_j,\, t_j)}\,\,.
  \end{split}
  \label{Eq:LW-values}
\end{equation}

While the age $t$, the metallicity $Z$ and the mass $m$ of the particles cover continuous ranges, the luminosity weights $w$ depend on the template's $ML_{\rm {SSP}}$ which has a discrete sampling in age and metallicity.

Because the particles get assigned a spectrum from the template, the fiber spectrum is formed by a combination of discrete ages and metallicities. To analyze how the intrisic ages and metallicities are affected by this discretizing step, an additional set of control properties is generated. These are calculated as Eq.\,\ref{Eq:LW-values}, but  replacing the ages and metallicities by the assigned values of the SSP template.

\subsection{Building MaNGA-like datacubes}\label{Subsec:mocks-obs}

To realistically emulate MaNGA observations, we include a series of instrumental effects that are detailed in this section. We first reproduce the fiber bundle dithering scheme (Sect.\,\ref{Subsec:mocks-FOV}) to obtain the spatial sampling and resolution of MaNGA. The spectrum corresponding to each fiber is constructed as in the previous section (Sect.\,\ref{Subsec:mocks-spec}). Noise is added to the spectra. Finally, the fiber spectra are re-sampled to MaNGA's spatial grid format.

The spectra observed in MaNGA are affected by noise introduced by the detector, of a Poisson distribution nature. Because the spectra are sky-subtracted to the Poisson-limit level, the noise is ideally distributed following a Gaussian distribution \citep{MaNGA-Law-2016}. We follow the recipe of \cite{Hector-mock} to emulate BOSS spectrograph noise which mimics the detector's behaviour by increasing the noise toward the edges of the wavelength range, with a plateau value $F_0$. As MaNGA observations target a signal to noise ratio (SNR) of 5-10 at the outskirts of the galaxies (at $1-2 R_{\rm e}$) \citep{MaNGA-Bundy}, we fix $F_0$ to obtain a $\rm{SNR}\sim5$ in the $r$-band at $2\, R_{\rm e}$. Because the noise is added to the fiber spectra before computing the spatial grid, the value of $F_0$ is estimated from the hexagonal ring of fibers with radius closest to 2\,$R_{\rm e}$ (see Fig.\,\ref{fig:ifu-rings}). This produces a wider variety of SNR in the mocks, which is more similar to observations.

The stacked spectra corresponding to each galaxy are then used to reconstruct the spectral cube. MaNGA's spatial grid is formed by $0.5 \times 0.5\,{\rm arcsec}^2$ pixels. The contribution of a fiber to the pixels in the grid is weighted by the distance $r_{p,i}$ of the pixel $p$ to the centre of the fiber $i$. If $r_{p,i}>1.6\arcsec$ the contribution of the fiber to the pixel is zero. Otherwise, the weights are defined as the following expression

\begin{equation}
    \omega_{p,i} = \exp\left(-\frac{r_{p,i}\,^2}{2\,\sigma_{\rm rec}^2}\right)/\omega_{\rm T}\,\,,
\end{equation}

where $\sigma_{\rm rec}=0.7\arcsec$ \citep{MaNGA-Law-2016} and $\omega_{\rm T}$ is a normalizing factor to keep the total flux constant. 

The mock datacubes have the format detailed in \cite{pipe3d-10000} to be analyzed by \textsc{pyPipe3D} \citep{pyPipe3D}, namely a FITS file comprising three extensions: the flux datacube in units of $10^{-16}\,{\rm erg\,s^{-1}\,\AA\,cm^{-2}}$ per spaxel, a second datacube with the flux error and a final extension comprising a datacube with a mask for bad pixels, all with a linear fixed step in wavelength. 

The intrinsic and assigned values of the galaxies that were saved per fiber for control (Sect.\,\ref{Subsec:mocks-spec}) are now spatially recombined in the same way as the mock datacubes. This results in a set of 2D maps with the instrinsic and assigned stellar properties of the simulated galaxies with the same spatial resolution as the datacubes. We refer to the maps generated from the continuous properties of the galaxies as the \emph{intrinsic} maps, while those formed with the discrete ages and metallicities from the template are the \emph{assigned} maps.

\subsection{From spectra to stellar population maps}\label{Sec:SSP-ana}

The post-processing of the mock datacubes is performed with \textsc{pyPipe3D} \citep{pyPipe3D}, the \textsc{Python} implementation of the \textsc{Pipe3D} code (\citealt{pipe3D-1}, \citealt{pipe3D-2}). The goal of this programme is to disentangle the contributions to the emission produced by the different stellar populations and extract the emission line information from galaxy spectra. 

For this purpose, it is necessary to separate the kinematic and dust extinction effects in the observed spectra and remove emission lines before finding the best-fitting SSPs. Therefore, the first step this software performs on given galaxy spectra is a non-linear fit to determine the extinction ($A^*_V$) and the stellar kinematic ($\sigma_*$ and $z$) parameters. A parametric fit is then used to approximate strong emission lines. Once the emission line contribution is calculated, it is subtracted from the original spectra to perform the stellar population linear fit. 

The SSP fit consists in finding the best-fitting linear combination of SSP spectra (positive contributions only). As the spectra generally have lower SNR toward the edges, a binning scheme is necessary to extract the SSP information from these regions. In particular, \textsc{Pipe3D} uses a Continuous plus S/N binning algorithm to define how adjacent bins are going to be co-added. As this scheme sets a goal S/N at the same time as it requires continuity in the surface brightness, the shape of the resulting spatial bins follows the isophotes, which typically have an arc shape rather than the square or round bins formed with the Voronoi binning scheme.

Once the stellar spectra have been decomposed into a set of SSPs, the code saves the spatially resolved star formation histories of the galaxy, as well as its maps of stellar age, metallicities and kinematics (see Fig.\,\ref{fig:spec_at_R}, recovered maps on the right hand-side). Other outputs of \textsc{pyPipe3D}'s analysis include the emission line properties and stellar absorption indices.

\section{Results}\label{Sec:overview}

We evaluate the performance of the complete pipeline by comparing the stellar population and kinematic maps produced during the mocking process (TNG50 intrinsic, assigned and recovered maps) and latter compare the simulated sample with the observed one (TNG50-recovered and MaNGA-recovered).  While the maps of the simulated galaxies were obtained as described in Sect.\,\ref{Sec:mocks}, those of the observed galaxies correspond to the latest \textsc{pyPipe3D} analysis release for $10,000$ MaNGA galaxies\footnote{\url{http://ifs.astroscu.unam.mx/MaNGA/Pipe3D_v3_1_1/index.html}} \citep{pipe3d-10000}. In both cases the fitting code and the SSP template used for the analysis are the same. The maps considered in this work are the $V$-band reconstructed image, the luminosity-weighted age and metallicity, LOS velocity and dispersion. First, we study the effect of discretizing in age and metallicity when associating stellar particles to a spectrum in the \texttt{MaStar\_sLOG} template. Second, we analyze how well the parameters of the simulated galaxies are recovered from the spectral fitting by comparing with a set of analogue stellar population maps calculated directly from the particle information of each simulated galaxy (assigned maps). Third, we compare the SSP maps derived from our mock datacubes with those derived from MaNGA observations.

For this section, all the stellar population maps have been masked when the median intensity had $S/N<3$. Because the mock datacubes have been produced with the largest IFU bundle of $127$ fibers, an additional mask was applied when their corresponding IFU bundle size is smaller (see Sect.\,\ref{Subsec:mocks-FOV}). In this case, the allocated IFU size is the same as its observed counterpart.

\subsection{Discretization effect in age and metallicity}\label{Subsec:discretize}

The simulations produce stellar particles with ages and metallicities that span in continuous ranges (constrained only by numerical limits). When the mock datacubes are produced, the particles are assigned to a limited number of SSPs ($273$) with discrete distributions in age and metallicity. The way the SSP grid is defined will determine how well the continuous values are represented by the template. 

\begin{figure}
    \centering
    \includegraphics[width=0.9\hsize]{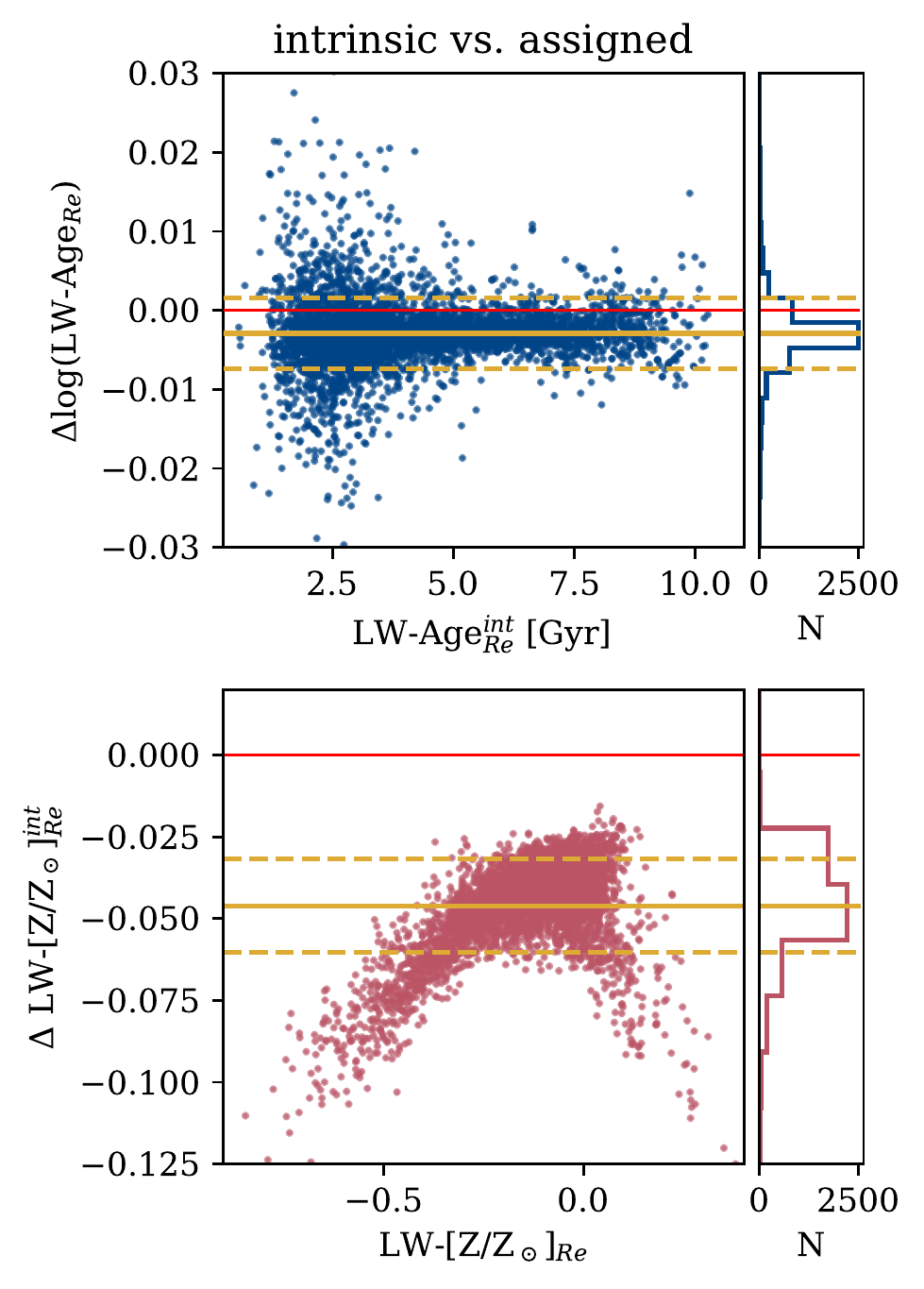}
    \caption{Difference between the intrinsic and assigned mean LW-age/metallicity (top/bottom) within $1R_{\rm e}$ per galaxy (dots), calculated from the respective 2D maps. The yellow solid/dashed lines indicate mean/standard deviation of the differences of the complete set of galaxies. While ages are slightly younger after being assigned, metallicities show a greater negative deviation from the intrinsic values.}
    \label{fig:int-assign_bias}
\end{figure}

We compare the intrinsic LW age and metallicity of the simulated galaxies before and after being assigned an SSP from the template. For this, we calculate the mean LW age/metallicity within $1R_{\rm e}$ from both the intrinsic and assigned LW age/metallicity maps per simulated galaxy. The difference between these values is shown in Fig.\,\ref{fig:int-assign_bias} (assigned minus intrinsic in log-scale). The age differences spread around zero with a larger scatter toward young ages (ages$<5$\,Gyr). The mean of all differences is $\Delta{\rm log}({\rm Age}_{R{\rm e}})<0.003$\,dex, which is smaller than the standard deviation of $0.004$\,dex. 

After assignation, the LW metallicity is systematically lower than the intrinsic one. Furthermore, the assigned values tend to deviate more from the intrinsic values at low ($[Z/Z_{\odot}]<-0.3$) and high ($[Z/Z_{\odot}]>0$) metallicities, reaching maximimum differences of $\sim0.1$\,dex. The greater deviation toward high metallicities could be related to TNG50 galaxies generally having particles with metallicities greater than the template's highest value ($[Z/Z_{\odot}]>0.372$), being more frequent at higher masses (see Appendix\,\ref{Ap:met-excess}). These particles' metallicities are underestimated when discretized, resulting on an average lower metallicity of the galaxy. Within the range where the particles' metallicity does overlap with the template's range of metallicity, the discretization could still produce, on average, differences with the intrinsic values. This could be because the intrinsic values distributions of the simulated galaxies may not be well represented by the sampling of the template, or the interpolation method should be revised. 

\subsection{Recovering physical properties with \textsc{pyPipe3D}}\label{Subsec:pipe3d-bias}

Determining age, metallicity and extinction from a spectrum is a highly degenerate problem. Even in the most favourable case where the input spectrum is a linear combination of spectra from the SSP template, recovering ages and metallicities through template fitting is a challenging task. Therefore, to see how well these properties are disentangled by \textsc{pyPipe3D}, we compare the stellar population maps recovered with \textsc{pyPipe3D} (recovered) with a set of maps that were produced directly from the stellar particle properties after being assigned the nearest ages and metallicities from the template's discrete grid (assigned maps, see Sect.\,\ref{Subsec:mocks-spec}). The latter are produced with the same spatial properties as the mock datacubes so that the resolution is comparable. Additionally, the age and metallicity maps are luminosity-weighted and averaged in log-scale, as those recovered by \textsc{pyPipe3D}. 

Qualitatively, besides differences due to the binning in the recovered maps, we find that the spatially resolved properties are properly recovered, as seen in the examples in Fig.\,\ref{fig:spec_at_R} (right-hand side panels).

\begin{figure}
    \centering
    \includegraphics[width=\hsize]{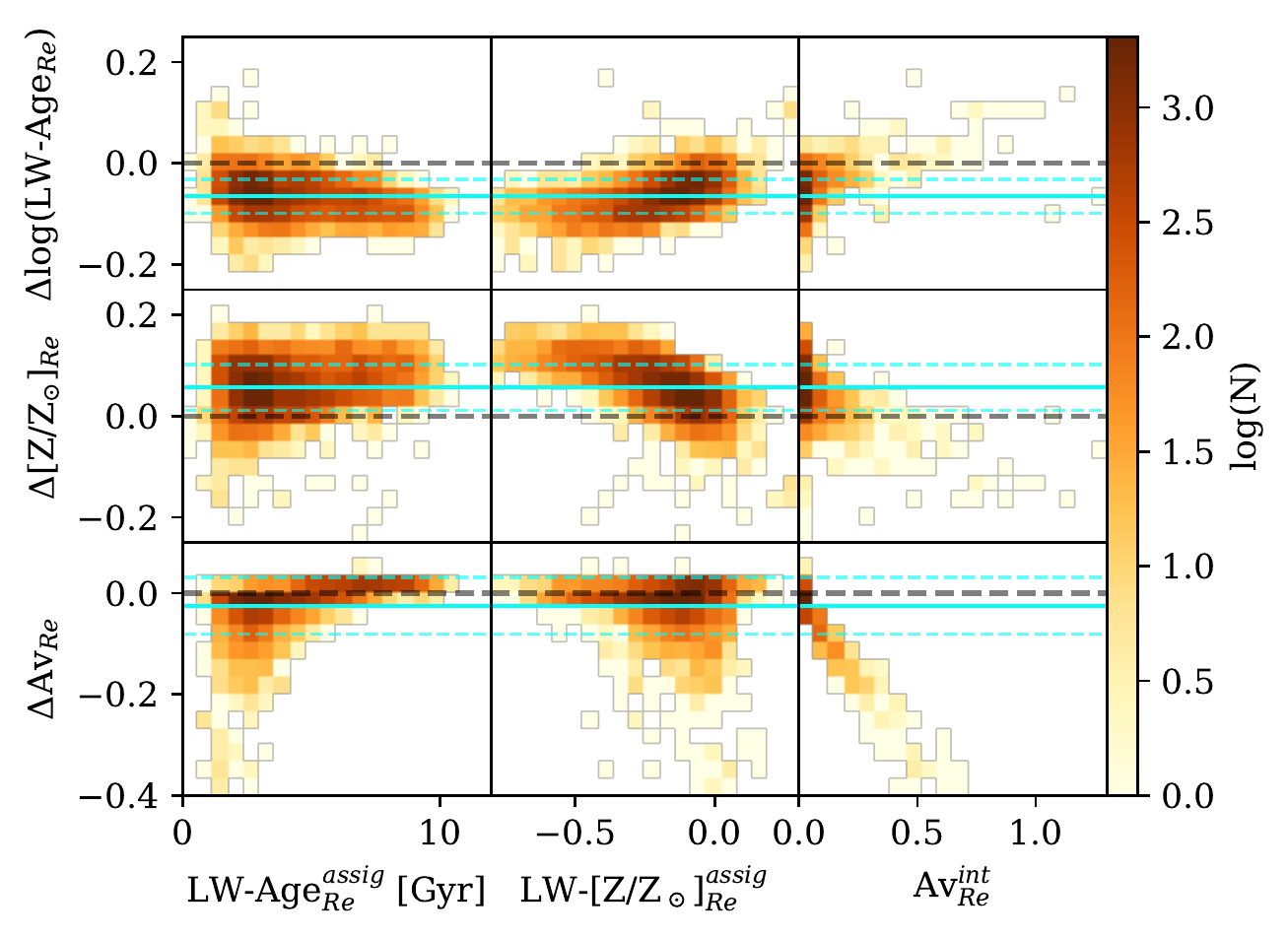}
    \caption{Differences in age, metallicity and extinction recovered by \textsc{pyPipe3D}. The difference $\Delta$ is calculated as \emph{recovered-assigned} for age and metallicity [dex], and \emph{recovered-intrinsic} for $Av$. Panels show the number density of galaxies of the differences across each parameter. The solid/dashed cyan lines show the mean/standard deviation of the differences. A age-metallicity degeneracy is seen, as \textsc{pyPipe3D} tends to recover systematically younger ages and higher metallicities than expected. However, the recovered values are within the expected errors of $0.1$\,dex. Extinction is increasingly underestimated as its value grows.}
    \label{fig:pipe3d_bias}
\end{figure}

As the degeneracy in the determination of ages and metallicities is case-sensitive, it is worth analyzing how the age and metallicity estimations depend on the galaxy properties. We calculate the relative difference between the average values within $1\,R_{\rm e}$ of the recovered and assigned maps for each galaxy. In Fig.\,\ref{fig:pipe3d_bias} we see that the estimated values of extinction tend to be lower than the real ones as their values grow. Although this means that the recovered spectra are redder than the original ones, this is not evidently compensated by an older or metal-rich estimation. There is, however, a degeneracy between age and metallicity. The ages estimated from the recovered data tend to be systematically lower than those of the assigned maps ($-0.05$\,dex on average) and the metallicities tend to be estimated with higher values ($+0.05$\,dex on average). This effect is more notorious toward low metallicities, where the differences in ages and metallicity are around $0.1$\,dex. Although we find systematics in the recovered ages and metallicities, we highlight that the deviation from the assigned values is within the expected range of $0.1$\,dex (\citealt{gsd156}, \citeyear{cid-fernandes-2014}, \citealt{pipe3D-1}, \citealt{pyPipe3D}).

\subsection{Comparing simulations to observations} \label{Subsec:results-recovered}

We now confront the recovered SSP maps of simulated galaxies with their observed counterparts. Given the size of the sample, such detailed analysis cannot be done for all the galaxies. Therefore we compare integrated properties of the whole sample in the following subsections.

\subsubsection{Individual stellar maps}

\begin{figure*}
    \centering
    \includegraphics[width=\hsize]{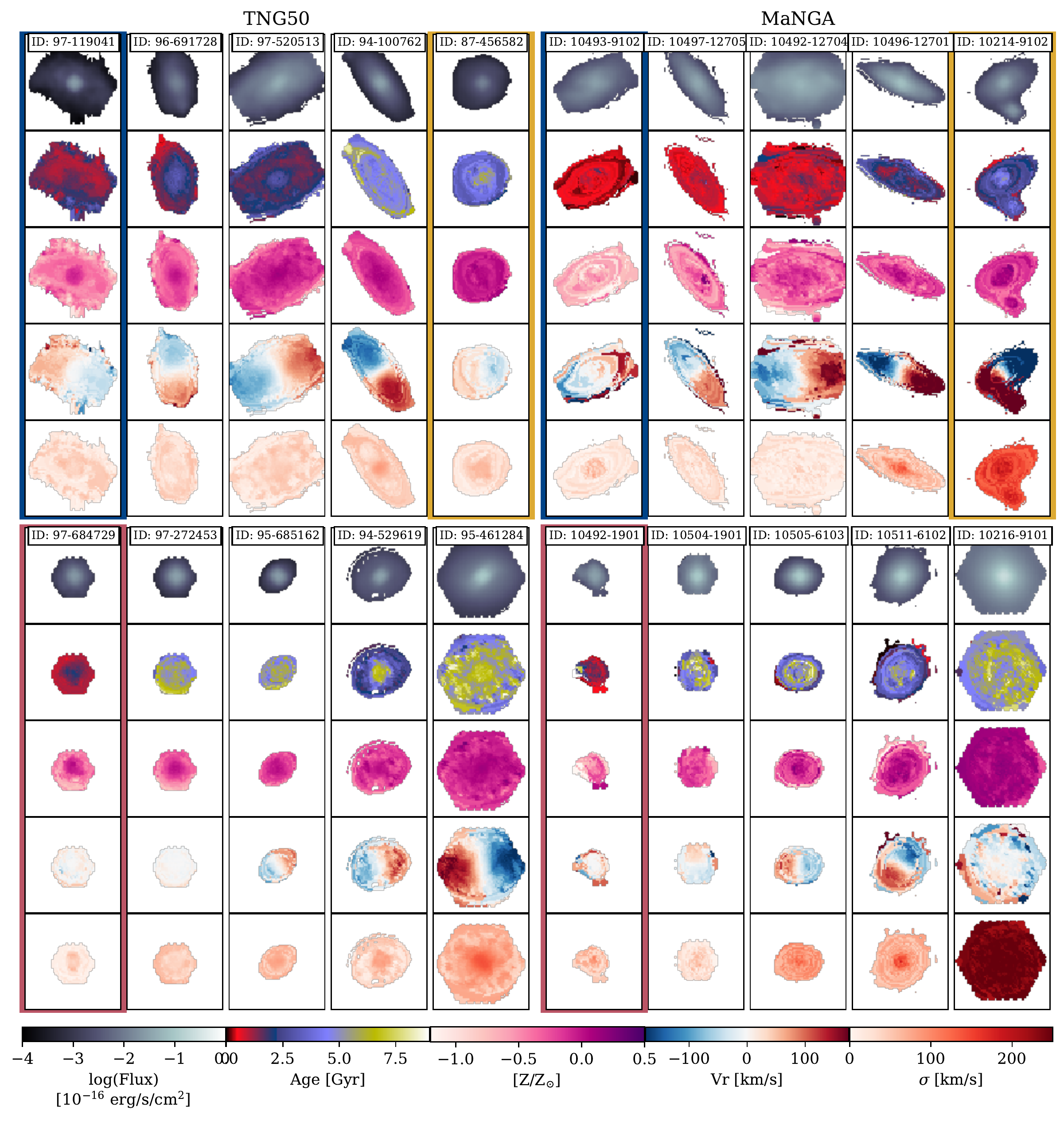}
    \caption{Left-hand side panels show ten TNG50 recovered stellar population and kinematic maps. From top to bottom in each sub-panel: the $V$-band reconstructed image, LW-age, LW-metallicity, LOS velocity and velocity dispersion maps. The right-hand side panels show the MaNGA counterparts of each of the left-hand side galaxies. These examples were chosen to represent ten regions of the $M_*-R_{\rm e}$: five mass bins (delimited by $10^{9.25}, 10^{9.75}, 10^{10.25},$ and $10^{10.75}\,M_{\odot}$) that correspond to the columns, with increasing mass from left to right, and two $R_{\rm e}{\rm [kpc]}$ regions per mass bin (rows, increasing size upward).}
    \label{fig:maps-example}
\end{figure*}

 To assess how well the mocks reproduce the observational features in the data, we show the stellar population and kinematic maps obtained for ten example galaxies as compared with their MaNGA-observed counterparts (Fig.\,\ref{fig:maps-example}). These galaxies have been chosen to represent ten regions of the $M_*-R_{\rm e}$ plane. For this purpose, we have defined five mass bins (separated by the following mass limits:$10^{9.5}, 10^{10}, 10^{10.5}$ and $10^{11}\,M_{\odot}$) and each of them subdivided in two sizes (first and last $50^{th}-$percentile in $R_{\rm e}\,{\rm [kpc]}$ within the bin). We then choose the galaxies with the most similar mass and size to the median values in each of the ten bins. Because there are two galaxy sizes for the same mass, we refer to the galaxies with larger $R_{\rm e}$ as extended while the smaller ones are compact.

 Mocks realistically reproduce the observational/pipeline effects seen in the observations including spatial resolution and binning. The noise level is adequate to recover a segmentation that is comparable to observations, as determined by the $S/N<3$ clipping in both cases (Fig.\,\ref{fig:maps-example}).

 While only ten example galaxies per mock/observed sample are shown in Fig.\,\ref{fig:maps-example}, these were chosen to have representative masses and sizes of different regions of the $M_*-R_{\rm e}$ plane. We see that the galaxies are older, more metal-rich and exhibit less rotation as they are more compact and massive. The extended simulated low- and intermediate-mass examples have a distinctive central component and show negative gradients in age and metallicity, while the observed counterparts show flatter distributions and a more prominent disc component. At high mass, the compact simulated examples are fast-rotating, while the observed galaxies have higher velocity dispersion and disrupted LOS-velocity maps, corresponding to a typical slow-rotating galaxy (bottom) and an interaction with a nearby galaxy (top).

\subsubsection{Integrated properties}

\begin{figure*}
    \centering
    \includegraphics[width=\hsize]{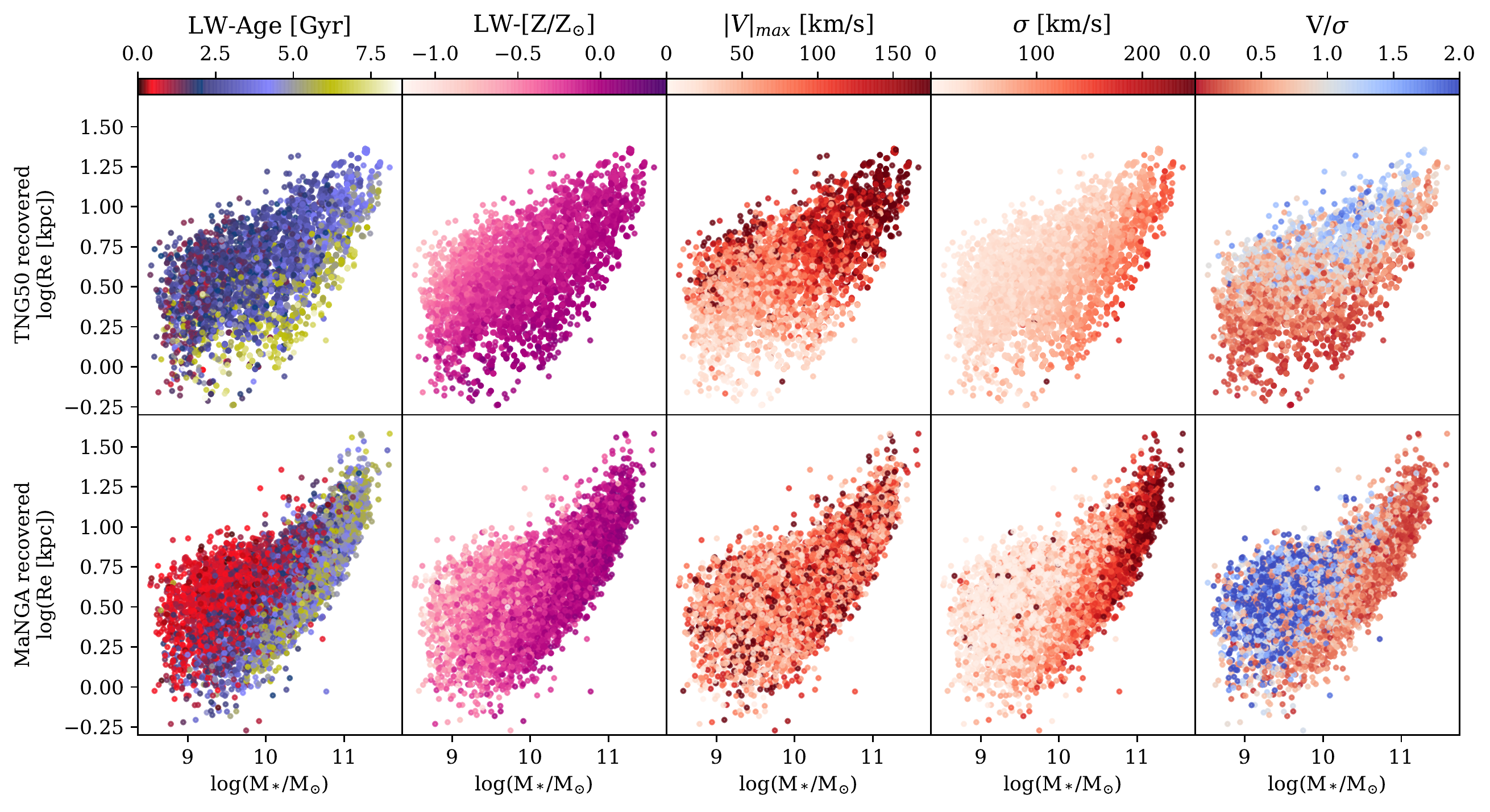}
    \caption{Top row panels show TNG50 galaxies in the $M_*-R_{\rm e}$ plane, while the bottom row shows the MaNGA counterparts. Galaxies are colour-coded from left to right by their mean LW-age, LW-metallicity, LOS maximum velocity, dispersion and projected $V/\sigma$ calculated within $1\,R_{\rm e}$ from the recovered maps. The trends seen in the observed sample are replicated in the simulated sample with a greater dependence with galaxy size rather than mass. Although there is discrepancy between the maximum values in dispersion and minimum values in age, the galaxies with such values lie in the analogue regions of the plane.}
    \label{fig:mean-prop}
\end{figure*}

To confront the global properties of the observed sample with those of the mocked one, we analyze how the mean values of age, metallicity, LOS maximum velocity, projected $V/\sigma$ and dispersion distribute across the $M_*-R_{\rm e}$ plane. All values are calculated within $1\,R_{\rm e}$. While the projected $V/\sigma$ is calculated as in \cite{Cappellari-vsig}, the values obtained can differ from other works since this parameter is sensitive to the definition of the reference systemic velocity of the galaxy and the spaxel binning, which in our case is CS-binning instead of Voronoi. However, the values shown in Fig.\,\ref{fig:mean-prop} are directly comparable as they have been calculated in a consistent manner for observed and simulated galaxies. 

For the observed sample, older and more metal-rich galaxies lie in the lower and more massive side of the plane, as seen in Fig.\,\ref{fig:mean-prop} (top panels). These main trends are replicated in the matched TNG50 sample (bottom panels). However, the age, metallicity and $V/\sigma$ of the simulated sample have a stronger dependency on the size of the galaxy than those of the observed sample, which show a clearer correlation with stellar mass. In contrast, the LOS maximum velocity is strongly correlated with stellar mass in the simulations, with the highest values in the most massive end. This is consistent with the lack of simulated galaxies with low $V/\sigma$ in this region, while the lack of galaxies with high $V/\sigma$ at low masses could be due to slightly higher values of dispersion than in observations. 

As expected, higher velocity dispersion values are found in the most massive end. However, the highest velocity dispersion of the simulated galaxies is limited to $\lesssim150\,{\rm km/s}$ while observed galaxies reach values of $250\,{\rm km/s}$. Similarly, the youngest galaxies of both samples lie in the low- to intermediate-mass and most extended regions of the $M_*-R_{\rm e}$ plane, but here the most frequent age for the simulated galaxies is around $2\,{\rm Gyr}$, while for the MaNGA galaxies this value is $1\,{\rm Gyr}$ younger.

\subsubsection{Gradients}\label{Subsec:gradients}

\begin{figure*}
    \centering
    \includegraphics[width=\hsize]{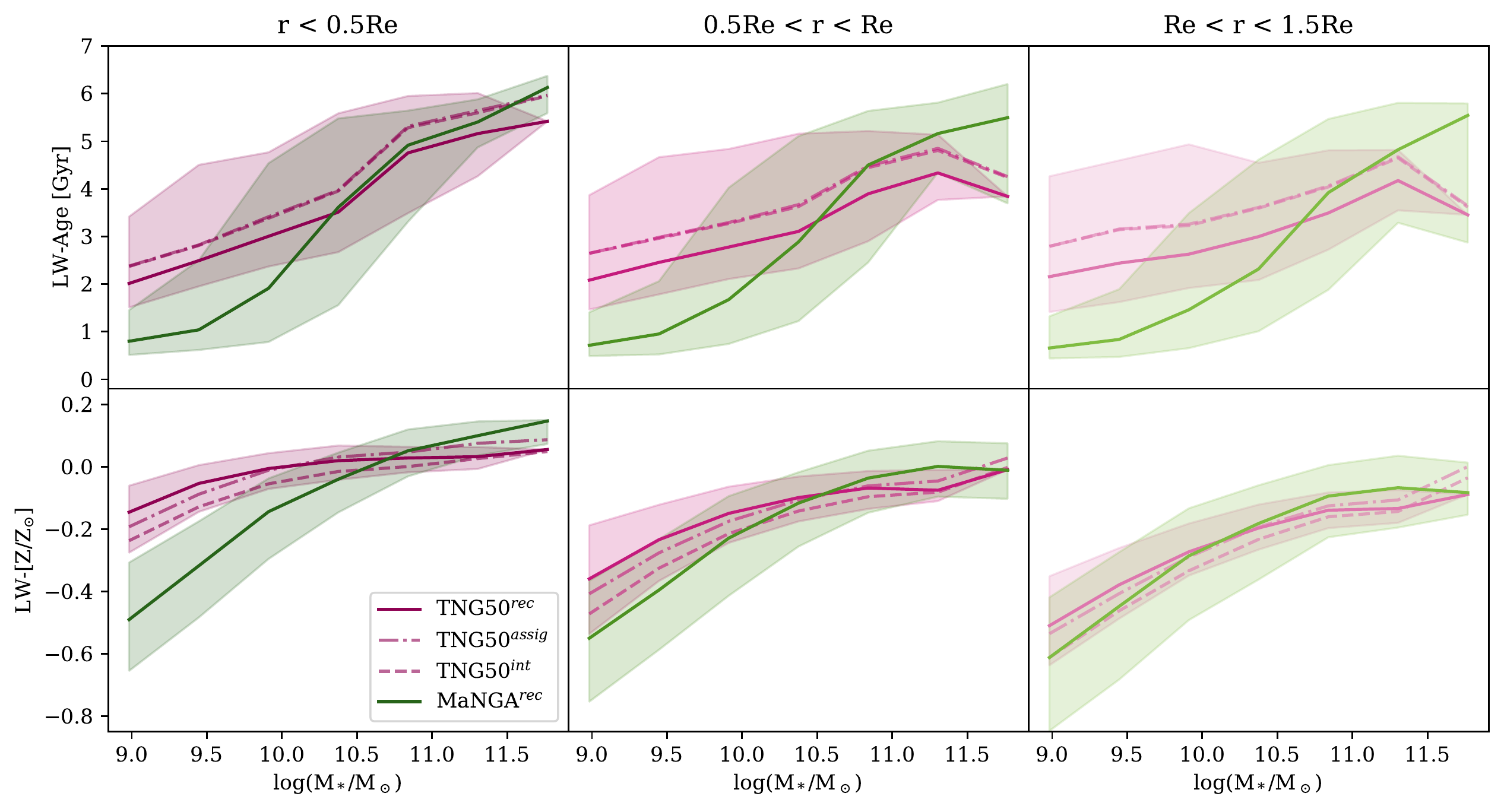}
    \caption{Median luminosity-weighted age (top panels) and metallicity (bottom panels) per mass bin at increasing radii for MaNGA galaxies (green) and TNG50 fully-mocked with \textsc{pyPipe3D} (solid magenta line). The median values for the TNG50 intrinsic/assigned maps are shown with dashed/dot-dashed lines to compare the median shift produced by the mocking process. The shaded areas span from the $20^{\rm th}$ to the $80^{\rm th}$ percentile. }
    \label{fig:agemet_mas_grd}
\end{figure*}

To analyze how the age and metallicity of the observed sample and the mock counterpart vary across the galaxies, we consider the mean values of the LW-maps in three annuli: within $0.5\,R_{\rm e}$, between $0.5\,R_{\rm e}$ and $1R_{\rm e}$, from $1R_{\rm e}$ to $1.5\,R_{\rm e}$. Then, we calculate the median values per mass bin for both samples.

The best match in age between observations and mocks is found within $0.5\,R_{\rm e}$ for high-mass galaxies ($10^{10.5}M_{\odot}<$), even though for lower-mass galaxies the discrepancy is of $1\,{\rm Gyr}$ (top row in Fig.\,\ref{fig:agemet_mas_grd}). Toward the outer regions of the galaxies, the discrepancy between observed and mocked increases. The mocks have a flatter distribution of ages across mass outside $0.5\,R_{\rm e}$, while observed galaxies have a steep increase of ages with mass. 

The best match in metallicity between mocks and observations is found in the outer regions, in contrast to what is seen in age. Compared to observations, simulations form more metal-rich bulges in low- to intermediate-mass galaxies.

As seen before, there is some degeneracy when recovering age and metallicity with \textsc{pyPipe3D}, however the discrepancy between observations and simulations is still evident when we compare with the values obtained from the particle maps (dashed lines in Fig.\,\ref{fig:agemet_mas_grd}). As pointed out in the previous section, \textsc{pyPipe3D} tends to predict lower ages than the real ones. This effect is greater in the outer regions of the galaxies, suggesting that the code's performance might be affected by low S/N.

\section{Reliability of the sample and caveats}\label{Sec:discussion}

The trends of the observed physical properties are recovered in our sample, although there is still a mismatch between the samples. We discuss here how the assumptions made when creating the mock datacubes could be the source of this discrepancy.

\subsection{Simulations}

The mock galaxies are older than those observed for MaNGA in the low- to intermediate-mass regime. This behaviour has been observed before by \cite{Dylan-TNG100} for TNG100 galaxies. The results between that work and ours might originate in the different samples of galaxies used in each case and/or a resolution difference, since the cosmological volumes of the TNG50 and TNG100 simulations differ.

\begin{figure}
    \centering
    \includegraphics[width=0.8\hsize]{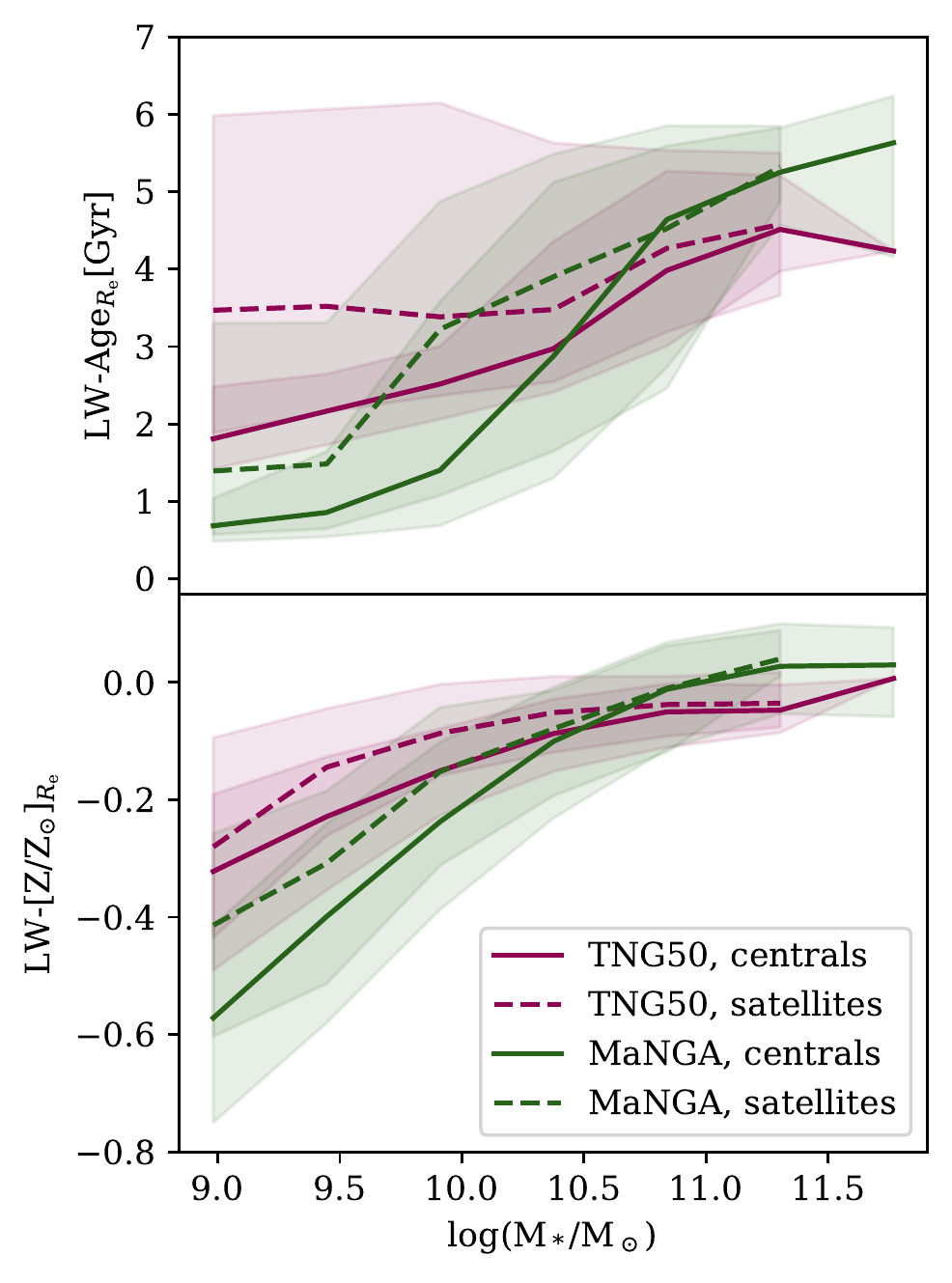}
    \caption{As Fig.\ref{fig:agemet_mas_grd}, but taking the mean age and metallicity per galaxy upto $1R_{\rm e}$. Solid/dashed lines represent the median age (top panel) and metallicity (bottom panel) per stellar mass bin of the central/satellite galaxies. TNG50 galaxies are plotted in magenta while MaNGA galaxies are plotted in green.}
    \label{fig:agemet_satcent}
\end{figure}

In the simulated sample, the compact galaxies have old ages ($>5$\,Gyr) and high metallicities ($[Z/Z_{\odot}]>0$) at low masses, which is not seen in the MaNGA sample. In contrast to more extended galaxies in the same mass regime, the simulated compact galaxies are old and, therefore, their age estimates should not be affected by strong emission lines. The difference between observed and simulated galaxies in this particular region of the $M_*-R_{\rm e}$ plane could be related to the larger fraction of satellite galaxies in MaNGIA ($55\%$) compared to MaNGA ($25\%$, based on the \citealt{yang-groups} group catalogue). To analyze the effect in satellite and central galaxies of the observed and simulated samples, we compute the median LW-ages and metallicities per mass bin as in Sect.\,\ref{Subsec:gradients} (age and metallicity per galaxy is calculated as the mean value within $1R_{\rm e}$ from the recovered maps), now separating satellites from centrals (We consider a galaxy as 'central' if it is not a satellite). The satellite galaxies are, on average, older and more metal-rich than central galaxies (see Fig.\,\ref{fig:agemet_satcent}), both in the observed and simulated sample. Older ages seen for  satellite galaxies are consistent with a gas stripping scenario (\citealt{Gunn-1972}, \citealt{Quilis-2000}), where a more massive nearby galaxy removes gas from the less massive satellite limiting its ability to form new stars. While central galaxies may accrete stars and gas with low metallicities resulting in a effective lower metallicity, the enhanced metallicity observed in satellite galaxies might be connected to "strangulation" processes, which operate more effectively in the low- to intermediate-mass regimes ($M_*<10^{10}M_{\odot}$) (\citealt{peng-2015}).

While the central and satellite galaxies have consistent behaviours in both observed and simulated sample when considered separately, there is still a discrepancy between the satellite/central simulated and observed galaxies. The simulated galaxies with stellar masses lower than $10^{10.5}M_{\odot}$ are older and more metal-rich than those observed, while at higher masses we find the inverse behaviour.

In the high-mass regime, the simulated galaxies show considerable rotation ($V/\sigma>1$, right-hand side top panel in Fig.\,\ref{fig:mean-prop}), which contrasts with the typically slow rotation in the observed galaxies. The behaviour of the massive simulated galaxies is compatible with the overpopulation of red spirals and late-type galaxies at high masses reported in \cite{vicente-statmorph} and  \cite{huertas-company-2019}, respectively. As these results were found for the TNG100 simulation, a different behaviour could be expected for TNG50. However, \cite{Donnari-2021b} show that TNG50 has a lower quenched fraction of galaxies at high masses than other TNG simulations (TNG100, TNG300), suggesting that the behaviour is still present in the highest resolution volume.

\subsection{Stellar population templates}

The determination of ages and metallicities from spectra relies heavily on the SSP template, which depends on the IMF, isochrones and stellar library assumed to build the template, as well as the sampling in age and metallicity of the template. The impact of the template choice in this work can be separated in two aspects. On the one hand, the generated spectra and the recovered maps of the mocks will depend on these choices. On the other hand, observed galaxies do not necessarily behave as the models. 

By selecting the same template to produce the mocks as to analyze the observed datacubes, the uncertainty introduced by the assumptions is minimized when recovering the stellar population properties of the simulated galaxies. In terms of age and metallicity estimations from the mock data, the most relevant effect of the template choice is in the luminosity-weighted maps as age and metallicity, since the $M/L$ ratio is given by the template. It is worth noticing that the IMF assumed in the template used in our mocks is different to the one used in the TNG simulation. However, as in \cite{Lorenza-mock}, we point out that this should not introduce large inconsistencies as both in the simulations and in the templates a universal IMF is assumed.

\begin{figure}
    \centering
    \includegraphics[width=0.8\hsize]{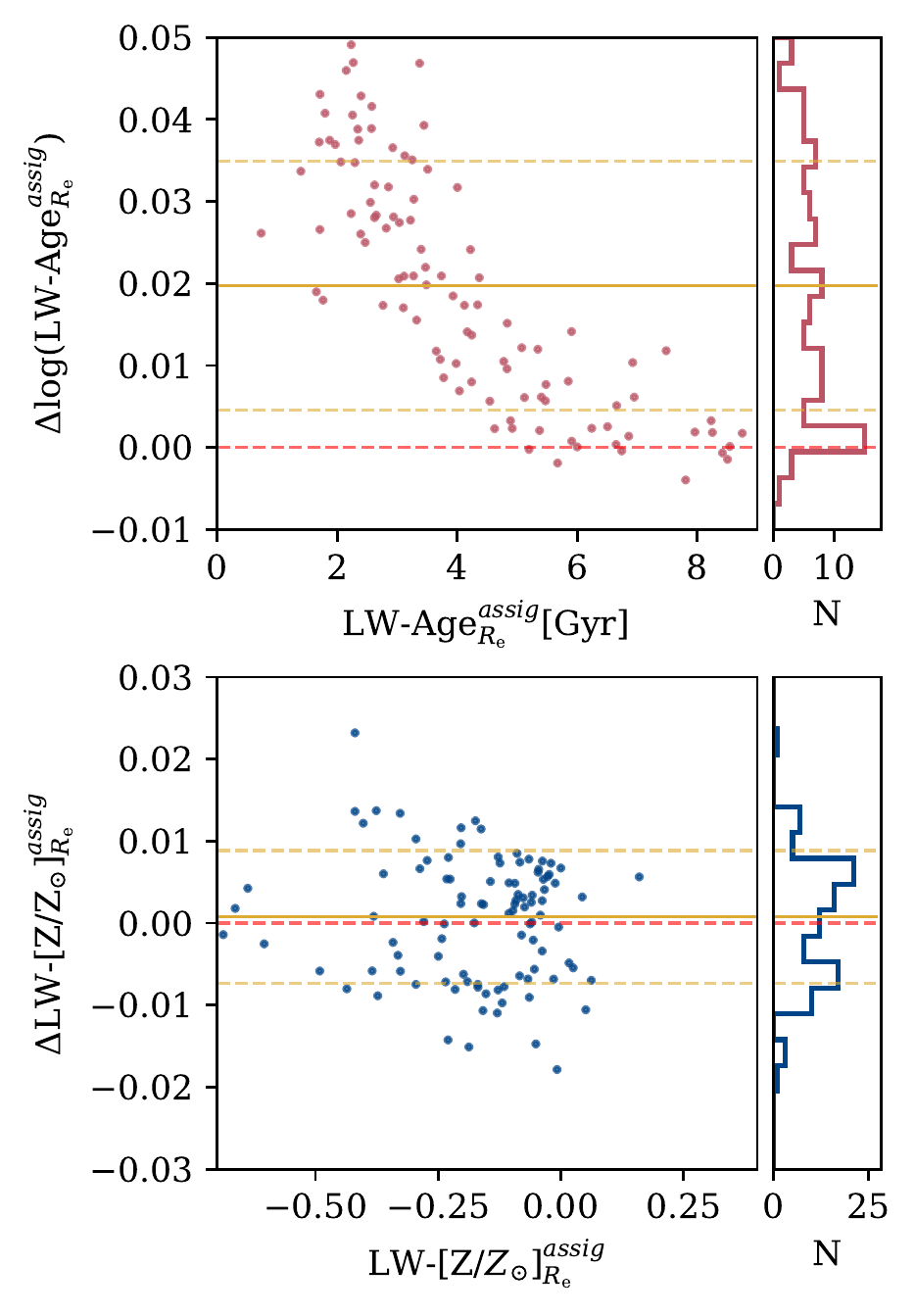}
    \caption{Difference in the LW-age (top) and metallicity (bottom) obtained with {\tt MaStar\_MA19} and {\tt MaStar\_sLOG} templates. The properties were calculated as the mean values within $1R_{\rm e}$ from the assigned maps. While the metallicities are comparable, the assigned ages tend to be older when using the {\tt MaStar\_MA19} template.}
    \label{fig:template_bias}
\end{figure}

For comparison, we produce a reduced sample of $100$ mock assigned LW-maps using the \cite{Maraston-mastar} template ({\tt MaStar\_MA19} from now on) and compare with the assigned maps produced with the {\tt MaStar\_sLOG} template, for the same $100$ galaxies. The {\tt MaStar\_MA19} template is based on \textsc{MaStar}, the same stellar library as the one used in this work. It combines \cite{cassisi-1977} and \cite{schaller-1992} isochrones, and it assumes a \cite{Kroupa} IMF. This IMF's positive slope for low masses ($<M_{\odot}$) and higher fraction of massive stars than Salpeter's IMF is more similar to \cite{Chabrier} the IMF, the one used in the TNG50 simulations. The difference between the two templates in LW-metallicity is negligible ($\Delta[Z/Z_{\odot}]<0.03$), and the difference in LW-age is only evident for young galaxies. These differences are smaller than the uncertainties of the parameter estimation during the post-processing of the data-cubes, as shown in Fig.\ref{fig:template_bias}. Although the {\tt MaStar\_MA19} template is based in the same stellar library as the one used in our work (Sect.\ref{Subsec:mocks-spec}), the differences observed can also be affected by the different sampling in age and metallicity in each template. Although this highlights the complexity in the age determination through SSP template fit, it also shows that the assumptions behind the different templates have a limited impact in the LW-ages and metallicities when generating the mocks. 

The age and metallicity sampling of the SSP template plays a role both when producing the mocks and in the retrieval of the stellar parameters with the spectral fit. Here we focus on the first effect. As seen in Sect.\,\ref{Subsec:discretize}, the sampling of the SSP produces a systematic shift in the average metallicity of galaxies, where intrinsic values are more metal-rich than those assigned (Fig.\,\ref{fig:agemet_mas_grd}, bottom row). On the one hand, this behaviour could be related to the metallicity range covered by the SSP template, as some stellar particles in TNG have metallicities that exceed the highest value in the template ($Z=0.4$, see Ap.\,\ref{Ap:met-excess}). On the other hand, the spacing of the metallicity template and/or the interpolation method used to assign an SSP spectrum to a stellar particle (nearest interpolation) may not properly represent the intrinsic properties of the particles, resulting in lower metallicities than expected.

Finally, the stellar population maps retrieved from the mock spectra are also affected by the internal degeneracy of the SSP template, since different combinations of spectra from the template could result in a similar added spectrum. This leads to age and metallicity estimates that deviate from the real ones. While age and metallicity estimates from real spectra are also affected by this, the effect can be quantified with the mocks because they were built as a linear combination of the spectra in the template. If there was no degeneracy, the combination of spectra should be perfectly retrieved.

Recovering ages and metallicities from mocks produced with the same SSP template as the one used in post-processing is an idealized scenario where the nature of the galaxies is fully known. While the uncertainty is minimized by choosing the same SSP to generate the mocks, the assumptions may not be identically valid for real galaxies. As the degeneracy when recovering ages and metallicities through template fitting is not negligible, this can lead to discrepancies between the samples. This could occur in particular for young stellar populations, where the variations in the spectra are more significant. 

We highlight that these properties in the mock sample recover the overall trends seen in the observed one. This is in spite of the complexity of age and metallicity determination with SSP fits on real spectra and the fact that the simulated sample has not been built to match MaNGA's age and metallicity distribution.

\subsection{Emission lines}

While interesting information is contained in emission lines, these are difficult to simulate reliably. Because studying the galaxies' line emission is not the final scope of this work, we produce spectra without it. Nonetheless, this choice motivates a further study of the effect of emission lines on the stellar population analysis. In Appendix\,\ref{Ap:emlines}, we show that strong emission lines can lead to lower estimates of metallicity and age. Although a detailed quantification of this effect is beyond the scope of this work, we find that it can be more important than the uncertainty in the retrieval of ages and metallicities when only considering the stellar component.

In our sample, this means that the match with observations in metallicity and age could be slightly improved by adding emission lines, as the discrepancy is stronger where the youngest galaxies lie (extended low- to intermediate-mass region in the $M_*-R_{\rm e}$). These galaxies are prone to be star-forming as they exhibit such young ages. Therefore, they should still have massive stars capable of ionizing the surrounding gas and producing emission lines.

\subsection{Comparison with other works}

Our work is the first to construct MaNGA-like mock data cubes of simulated galaxies from TNG50 that match the number and mass-size-redshift distribution of the observed MaNGA galaxy sample. This allows for extensive comparisons between simulated and observed galaxies connecting the observed properties of current-epoch galaxies to their formation and evolutionary history. Previous studies have only matched smaller subsamples of simulated galaxies to manga observations. We will discuss here the similarities and differences between the various approaches.

We consider the works by \cite{Hector-mock} (hereafter I-M), \cite{Becky-mock} (N21),\cite{Lorenza-mock} (N22), and \cite{Connor-mock} (B\&H). For a fast comparison, the characteristics of each of these works are condensed in Table\ref{Tab:other-works}.

 \begin{sidewaystable*}
  \caption[yutoi8u]{Comparison with similar works}
     \label{Tab:other-works}
 $$ 
     \begin{tabular}{lccccc}
        \hline
        \noalign{\smallskip}
        &  \citet{Hector-mock}  & \citet{Becky-mock}      & \citet{Lorenza-mock}       & \citet{Connor-mock} & Sarmiento+ (2022)\\
        \noalign{\smallskip}
        \hline
        \noalign{\smallskip}
        Simulation      & ART & \textsc{Gadget-3} & TNG50 & TNG50 & TNG50 \\
        &\citet{ART-1}&\citet{gadget-1}&\citet{TNG-FirstResults}&\citet{TNG-FirstResults}&\citet{TNG-FirstResults}\\
        &\citet{ART-2}&\citet{gadget-2}&\citet{TNG-FirstResults-2}&\citet{TNG-FirstResults-2}&\citet{TNG-FirstResults-2}\\
        \hline
        \noalign{\smallskip}
        Number of mocks & 8 & 1,500 & 2 & 893x4 & 10,000\\
        \hline
        \noalign{\smallskip}
        SSP template    & \texttt{GSD156} & \textsc{Starburst99} & MaStar (Mar) & - & \texttt{MaStar-sLog} \\
        &\citet{gsd156}&\citet{Leitherer_1999}&\citet{Maraston-mastar}&-&\citet{pipe3d-10000}\\
        \hline
        \noalign{\smallskip}
        Emission lines  & \textsc{Cloudy} & \textsc{MappingsIII} & \textsc{MappingsIII} & - & -* \\
        &\citet{cloudy}&\citet{mappingsIII}&\citet{mappingsIII}&&\\
        &&&\citet{Groves_2008}&&\\
        \hline
        \noalign{\smallskip}
        Radiative transfer & No & \textsc{Sunrise} & \textsc{Skirt} & - & No\\
        &&\citet{sunrise-1}&\citet{Skirt-1}&&\\
        &&\citet{sunrise-2}&\citet{Skirt-2}&&\\
        \hline
        \noalign{\smallskip}
        IFU fiber & Yes & No & No & Yes & Yes\\
        \hline
        \noalign{\smallskip}
        Data Analysis & \textsc{Pipe3D} & \texttt{ppxf} & \textsc{Firefly} & - & \textsc{pyPipe3D} \\
        &\citet{pipe3D-2}&\citet{ppfx-1}&\citet{firefly-1}&&\citet{pyPipe3D}\\
        &\citet{pipe3D-1}&\citet{ppfx-2}&\citet{firefly-2}&&\\
        \noalign{\smallskip}
        \hline
     \end{tabular}
 $$ 
\end{sidewaystable*}

N21 and I-M are based on numerical hydro-dynamical simulations of a single galaxy or a pair of merging galaxies. Because our goal is to generate a sample of galaxies that is statistically comparable to the MaNGA survey, a large enough sample is needed to cover a variety of galaxy types. In this sense, hydro-cosmological simulations are ideal, as they produce a palette of different galaxies in a realistic environment. Additionally, high resolution is required to capture the structures that can be resolved by MaNGA. We therefore use TNG50 simulations. The common choice with other works (B\&H and N22) stems from TNG50's good compromise between resolution, number of available galaxies and the fact that these are immersed in a cosmological context. 

To construct the spectra, we employ a simpler approach than N21 and N22, who run the radiative transfer (RT) codes \textsc{Sunrise} (\citealt{sunrise-1}, \citealt{sunrise-2}) and \textsc{Skirt} (\citealt{Skirt-1}, \citealt{Skirt-2}), respectively. Our spectra are the result of linearly combining the individual mass-weighted and doppler-shifted SSP spectra associated to each stellar particle, considering a simple dust screen model. RT codes use Monte-Carlo methods to emulate the physical processes associated to dust (scattering, absorption and emission) in the LOS of light-emitting sources. These codes perform a more complex treatment of dust as they consider a localized dust contribution and have been used to produce mock galaxy imaging (\citealt{vicente-statmorph}, \citealt{Torrey-2015}). However, as computation time scales with wavelength sampling, they are computationally very expensive when the aim is to produce high-resolution spectra ($R_{\rm \lambda}>2,000$). Furthermore, \cite{Zanisi-SDSSmocks} have shown that dust modelled by \textsc{Skirt} does not necessarily produce more realistic mocks, as photometric images produced using this code without including dust effects yield images that are more similar to observations. For these reasons, we stick to a simpler approach to dust emission that is less computationally demanding.

To build IFU datacubes, a complex pipeline is needed. While this work, B\&H and I-M reproduce MaNGA fiber observations to form the spatial grid of the datacubes, N21 and N22 mimic this effect by convolving a high-resolution datacube with a comparable PSF in the spatial axes. By generating the fiber spectra, the spatially correlated noise of the datacubes is reproduced, as the noise is introduced by the spectrograph's detector and propagated to the datacubes during the recombination of the fibers. While it is possible to combine the output of an RT code with the \textsc{RealSim} code used by B\&H to fully imitate MaNGA's pipeline, our approach (and I-M's) is considerably less time-consuming because the spectra are built per fiber, which translates to building up to $381$ (the largest IFU has $127$ and three dithering positions) spectra per datacube instead of producing a factor of $10$ more spectra to form the spatial grid that covers the IFU's FOV with MaNGA's spatial sampling ($70\times70$ pixels of $0.5\times 0.5\,{\rm arcsec}^2$). 

The stellar population templates used in each case are different, as they either use different stellar libraries or assume different stellar formation models (e.g. IMF, isochrones. See references in Table\,\ref{Tab:other-works}). As in our work, the approaches of I-M and N22 are intended to perform a stellar population analysis on the mock datacubes. Therefore, the SSP templates of choice are the same as the ones used to process the MaNGA observations with the different stellar population analysis pipelines. Both this work and N22 use SSPs based on MaStar library.

N22 and I-M simulate only emission lines due to star-burst regions. While N22 associate stellar particles younger than $4$\,Myr with the \textsc{MappingsIII} template (\citealt{mappingsIII}), I-M model emission lines assuming that the star-forming gas forms young stars with ages $2.6$\,Myr. Their emission spectral template was produced with the \textsc{Cloudy} code (\citealt{cloudy}). N21 not only includes emission from starburst regions, but also AGN activity. Although we do not include emission lines for the spectral fitting step, the emission lines are calculated as in I-M and saved in an extension of the datacube file. Besides the uncertainties of the modelling of such features, the emission lines have not been flux-calibrated. While other approaches use Mappings III, a specific young stellar populations template that includes emission lines associated with young stars, we adopt the same stellar template used for the SSP post-processing for the complete age range to avoid introducing further uncertainties in the SSP fit.

N21 find that kinematic maps can change significantly after emulating the whole mocking procedure by increasing the velocity dispersion measurement. This effect is boosted when including dust scattering and absorption in the \textsc{Sunrise} prescription. They find that the presence and treatment of AGN could also have an effect on the kinematic estimates. We do not report an increase in velocity dispersion after mocking.

\section{Conclusions}\label{Sec:conclusions}

We present a catalogue of $10^4$ TNG50 galaxies, called MaNGIA, that matches MaNGA's target selection while maximizing the selection of unique simulated galaxies. MaNGA-like datacubes were generated for each of the galaxies in the catalogue and these were post-processed to obtain stellar population and kinematic maps that are comparable to those in \cite{pipe3d-10000}. MaNGIA is the first sample of $10,000$ fully-mocked and post-processed data-cubes that mimic MaNGA's target selection. This sample not only reproduces the global physical properties of the observed galaxies, but it also emulates observational effects to produce realistic stellar population and kinematic maps. Because these maps were generated with the same pipeline, they are directly comparable with observations. 

This sample is ideal to connect the galaxies' observable properties with the physical processes occurring within them, as many of the latter cannot be directly observed (i.e. accretion history, black hole growth) but can be easily extracted from the simulations. Therefore, this dataset can be used to constrain the cause of observables and infer the history of the observed galaxies.

The sample is also well suited for Machine Learning applications given its large volume and realistic observational features. The sample, mock datacubes and derived maps will be made publicly available with the publication.

While only stellar mass, size and redshift were matched, the overall trends in age, metallicity and velocity dispersion are in good agreement with observations. The sample has been obtained with a one-to-one match to MaNGA galaxies, but the dataset presented in this work is intended to be used in a statistical manner as other key parameters have not been matched (e.g. star formation, morphology, colour).

A discrepancy between the observed and simulated samples found in age and metallicity for the low- to intermediate-mass galaxies is likely related to a combination of sample selection biases and pipeline effects. We report a greater proportion of satellite galaxies in MaNGIA than in obsertvations, which leads to an average increase of ages and metallicities in the lower-mass regime. However, the differences seen for the younger galaxies are more probably related to uncertainties in the SSP template dependence and the spectral analysis. On the one hand, the determination of ages in the young regime through template fitting has high uncertainties. On the other hand, the presence of emission lines in galaxy spectra can lead to underestimated ages and metallicities. Because we do not include emission lines in our mocks, it is reasonable to find a stronger discrepancy in galaxies with young stellar populations, as these galaxies have young stars and the gas to form them, and therefore produce emission lines. 

In the high-mass end, the discrepancy in kinematics is probably caused by a greater proportion of massive discs in the simulations than in observations. This is compatible with previous works (\citealt{vicente-statmorph}, \citealt{huertas-company-2019}, \citealt{Donnari-2021b}), although it requires further study.

Future work includes analyzing how the structures in the resolved stellar population and kinematic maps of the simulated galaxies reproduce the observed ones. 

Additionally, a more realistic emulation of emission lines is planned. This will be done by constraining the physical properties and geometries of the star forming gas cells.

\section{Data access and additional products}\label{Sec:data_access}

The data products will be available with the publication of this work. This includes the catalogue of TNG50 simulated galaxies, the RSS files from which the MaNGA-like datacubes are easily derivable, and the intrinsic, assigned and recovered stellar maps.

The code will be made fully available and designed in a block chain style as described in Sect.\,\ref{Sec:mocks} and shown in Fig.\,\ref{fig:method-scheme}. The intermediate files and final data products are:

\begin{itemize}
    \item {Particle file: files with particle information from the simulation. Position, velocities, mass, age, metallicities of stellar particles and gas cells in the FOV.}
    \item {RSS: FITS file comprising the stacked fiber spectra and their corresponding coordinates in the FOV. The running time for producing a RSS file scales with number of fibers/IFU used, number of stellar particles and gas cells, SSP age and metallicity sampling. Can take a few minutes ($<10\,{\rm min}$) up to 12 hours for massive galaxies. This process can be distributed in several CPUs to reduce computing times.}
    \item {MaNGA-like datacube: FITS file with the spectral datacube. It takes less than $15\,{\rm min}$ on $1$\,CPU to produce the datacube from the RSS file.}
    \item {\textsc{pyPipe3D} SSP output: FITS file comprising the 2D maps with recovered stellar properties.}
    \item {Control maps: 2D maps with the intrinsic and assigned stellar properties of the simulated galaxies. }
\end{itemize}

Together with the data, we release a set of masks to obtain the different bundles. Alternatively, the row-stacked spectra data is also available upon request and it is possible to recombine the fibers for the different bundles with the code available.

\section{Acknowledgments}

We thank Kevin Bundy for coming up with the name of the MaNGIA sample, as well as envisioning MaNGA in the first place.

R.S. thanks the Max-Planck-Institut für Astronomie for the hospitality during a visit.

We thank Ignacio Mart\'in Navarro for the interesting discussions on galaxy properties and how they are affected by different evolution channels.

We acknowledge financial support from the State Research Agency (AEI\-MCINN) of the Spanish Ministry of Science and Innovation under the grants "The structure and evolution of galaxies and their central regions" with reference PID2019\-105602GB\-I00/10.13039/501100011033 and "Galaxy Evolution with Artificial Intelligence" with reference PGC2018-100852-A-I00, from the ACIISI, Consejer\'{i}a de Econom\'{i}a, Conocimiento y Empleo del Gobierno de Canarias and the European Regional Development Fund (ERDF) under grants with reference PROID2021010044 and PROID2020010057, and from IAC projects P/300724 and P/301802, financed by the Ministry of Science and Innovation, through the State Budget and by the Canary Islands Department of Economy, Knowledge and Employment, through the Regional Budget of the Autonomous Community. 

R.S. thankfully acknowledges the technical expertise and assistance provided by the Spanish Supercomputing Network (Red Española de Supercomputaci\'on), as well as the computer resources used: the LaPalma Supercomputer, located at the Instituto de Astrof\'isica de Canarias.

H.I.M. acknowledges a support grant from the Joint Committee ESO-Government of Chile (ORP 028/2020).

The IllustrisTNG simulations were undertaken with computer time awarded by the Gauss Centre for Supercomputing (GCS) under GCS Large-Scale Projects GCS-ILLU and GCS-DWAR on the GCS share of the supercomputer Hazel Hen at the High Performance Computing centre Stuttgart (HLRS), as well as on the machines of the Max Planck Computing and Data Facility (MPCDF) in Garching, Germany.

\begin{appendix}
\onecolumn
\section{Mass and radius estimations}\label{Ap:r-fit}

 For every TNG50 galaxy, we can obtain the radius of a sphere containing half of its stellar mass $R_*^{3\rm D}$. This is a good estimate for the galaxy size, however, it may not be directly comparable with the half-light radius used in MaNGA. We therefore perform a fit that allows us to obtain a better approximation to the observed radius. We use the \cite{vicente-statmorph} catalogue, that lists photometric measurements of TNG50 and TNG100 galaxies with stellar masses greater than $10^{9.5}M_{\odot}$ calculated with the \textsc{Statmorph} code on mock images of such galaxies. We use the catalogue corresponding to snapshot 95, since it has the nearest redshift to MaNGA's mean $z=0.03$. As MaNGA galaxies span a small redshift range ($0.01<z<0.15$), the approximation holds for the whole range. From this catalogue we extract the semi-major axis of the ellipse containing half of the stellar light $R_{0.5}^{\rm Petro}$ and the circularized Petrosian radius $R_{\rm p}$ in the $r$-band. As these values are given in co-moving kpc per pixel, we  convert them to physical units, for SDSS resolution.

 We seek to find a relation between $R_*^{3 \rm D}$ and $R_{0.5}^{\rm Petro}$. For this purpose, we define five mass bins in which we perform a linear fit between the previously defined radii in log-scale physical units. We find that values for $R_{0.5}^{\rm Petro}$ are generally larger than those for $R_*^{3\rm D}$. The scatter of these fits are within $0.1-0.17$\,dex. As the galaxies considered in this work extend to a lower mass boundary than $10^{9.5}M_{\odot}$, the fit found for the lowest-mass bin is used for all galaxies with $M_*<10^{10}M_{\odot}$.
 
 To obtain a mass estimate similar to the one used in MaNGA, we first find an estimation for $R_{\rm p}$ following the previous procedure and later calculate the stellar mass within $2R_{\rm p}$.

\section{Isotropic views for repeated galaxies}\label{Ap:iso-views}

TNG50’s high resolution comes at the cost of reducing the size of the cosmological box, resulting in a reduced number of eligible galaxies. Additionally, each snapshot is volume-limited. Therefore, to match MaNGA's mass distribution and reproduce the MaNGA survey properties, galaxy repetition is needed (especially for the high-mass end) to reproduce the MaNGA survey properties. 

To define the number of repetitions needed we consider the following. While defining less repetitions leads to a larger variety of galaxies, more repetitions lead to a better match in terms of mass and size. An additional constraint comes into play when we consider how the views are defined. In general, we aim to sample galaxies with random orientations, and, if repeated, they should be observed from the most different angles as possible. As the simulated galaxies are randomly oriented in the cosmological box, we only need to define the orientation of the $n-$\,viewers to achieve this. Since galaxies generally have axial symmetries, we avoid selecting views placed in opposite directions as they would show the galaxy with the same inclination. Furthermore, we can measure how different views are by taking the absolute value of the cosine distance between the unit vectors that define the views, which is maximized if observers are oriented at $90^{\circ}$ of each other and minimized if they are parallel. To maximize the distance, the observers must be isotropically distributed.

\begin{figure*}
    \centering
    \includegraphics[width=\hsize]{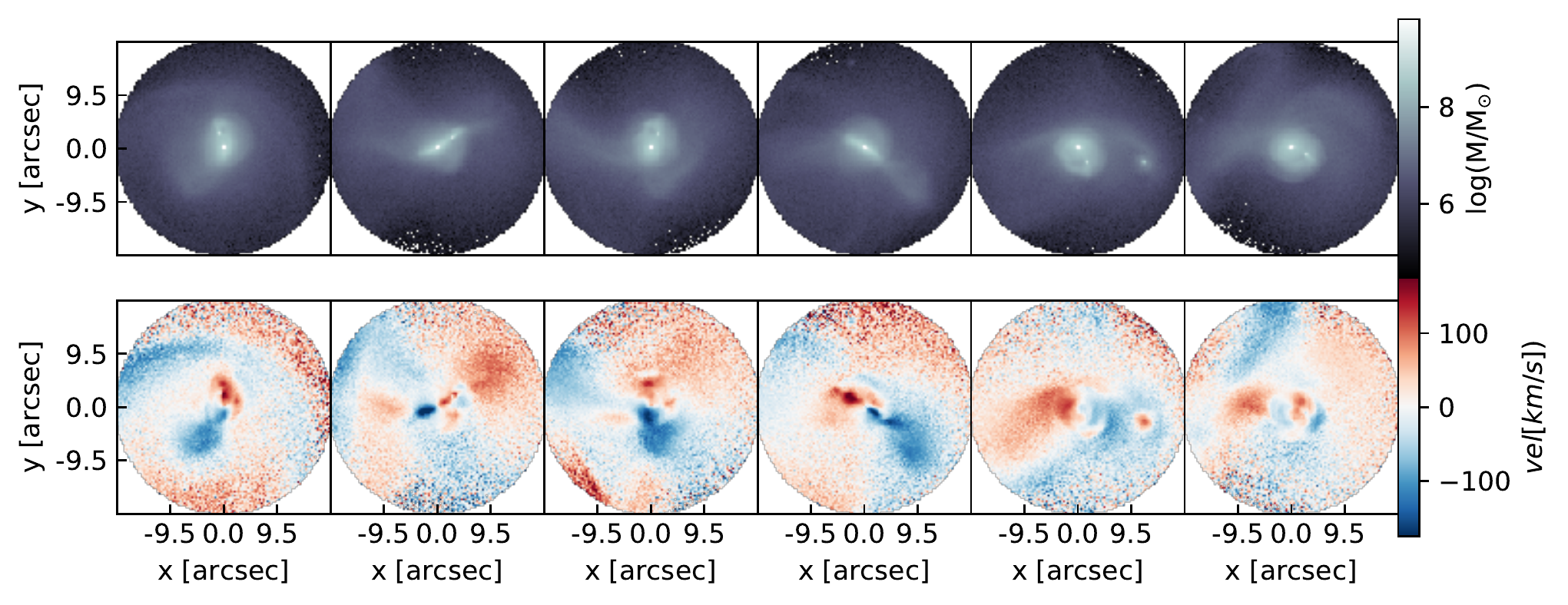}
    \caption{From left to right, six views of galaxy 377018 in snapshot 87. Top row shows the stellar mass per $0.25\arcsec \times 0.25\arcsec$ in the field of view. Bottom row shows the mean line-of-sight velocity of the stellar particles.}
    \label{fig:views6_example}
\end{figure*}

\begin{figure}
    \centering
    \includegraphics[width=0.5\hsize]{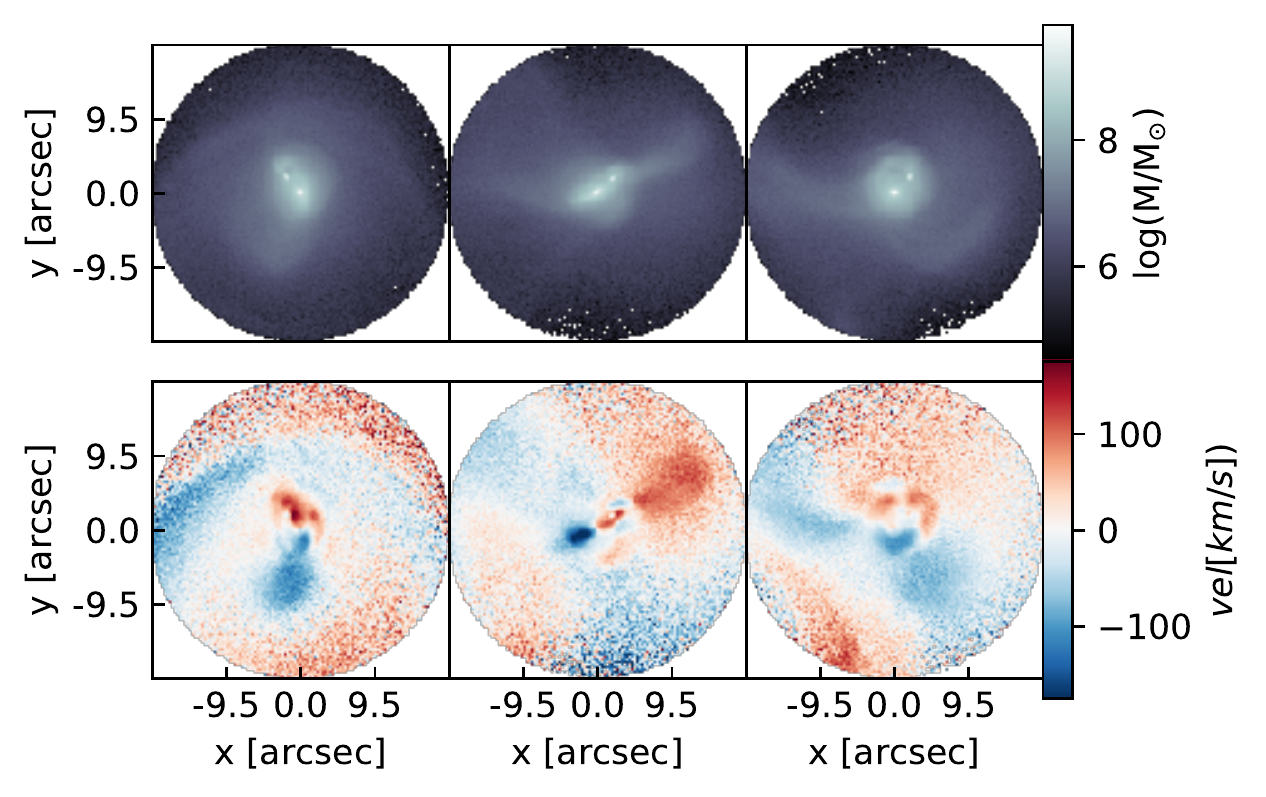}
    \caption{As Fig.\ref{fig:views6_example}, three views of galaxy 377018 snapshot 87. Note that the first three views in Fig.\ref{fig:views6_example} are chosen to be the most similar to the views in the $x$, $y$, $z$ directions.}
    \label{fig:views3_example}
\end{figure}

To obtain a set of six isotropic views (the maximum used in this work), we use the directions defined by the vertices of a regular polyhedron. In particular, we consider an icosahedron, a 3D shape with 20 faces and 12 vertices. We choose an icosahedron centred in the origin, oriented such that one of its vertices is in $(1,0,0)$ and another one in $(a, b, 0)$, where $a>0, b>0$. These vertices define the first and second views, respectively. The other four views are defined by the remaining vertices that have a positive first component. The minimum angular separation between two observers is $\alpha=63.4^{\circ}$. An example of how a galaxy would look with these six views is shown in Fig.\ref{fig:views6_example}, while the three standard views ($x$, $y$, $z$) for the same galaxy are shown in Fig.\ref{fig:views3_example}

\section{SSP analysis with emission lines}\label{Ap:emlines}

\begin{figure}
    \centering
    \includegraphics[width=0.8\hsize]{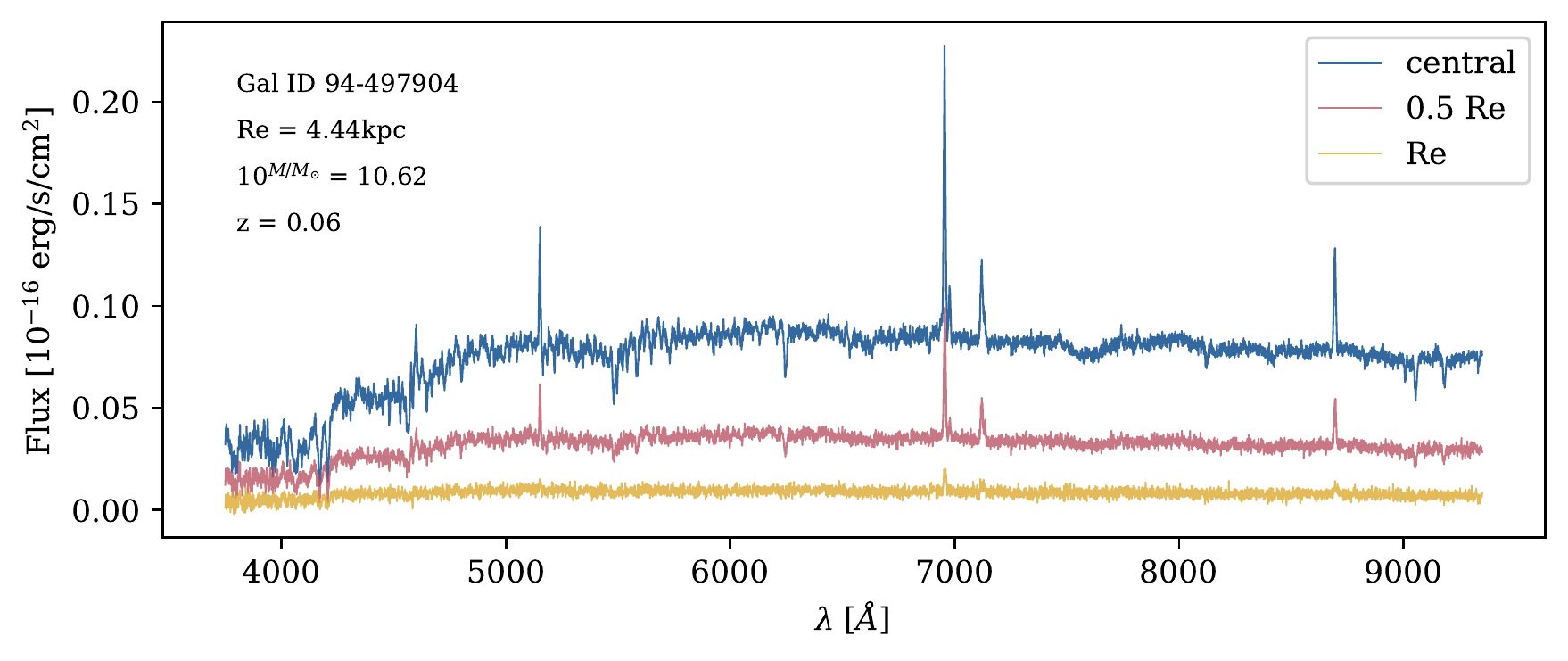}
    \caption{As Fig.\,\ref{fig:spec_at_R}, the mock spectra with emission lines of a simulated galaxy at its centre, $0.5\,R_{\rm e}$ and $1\,R_{\rm e}$.}
    \label{fig:EL-example}
\end{figure}

We study the impact of excluding emission lines (ELs) when retrieving stellar population contributions in the spectra. In this test we run \textsc{pyPipe3D} on a sub-sample of $100$ simulated galaxies twice, first adding the EL spectra and then without them. The sub-sample is selected to have a representative distribution in the $M_*-R_{\rm e}$ plane of the MaNGA sample. The difference between the recovered ages and metallicities in each case is then used as an indicator of the bias introduced in SSP fits by the ELs. We calculate the EL contribution as in \cite{Hector-mock} and downgrade its spectral resolution to MaNGA's typical values before adding it to the stellar continuum spectra. While the resulting spectra look realistic, the EL spectra may not be reliably modelled and, therefore, we do not pursue a further analysis of the ELs. An example of the spectra with the added contribution of the ELs is shown in Fig.\,\ref{fig:EL-example}.

\begin{figure}
    \centering
    \includegraphics[width=0.8\hsize]{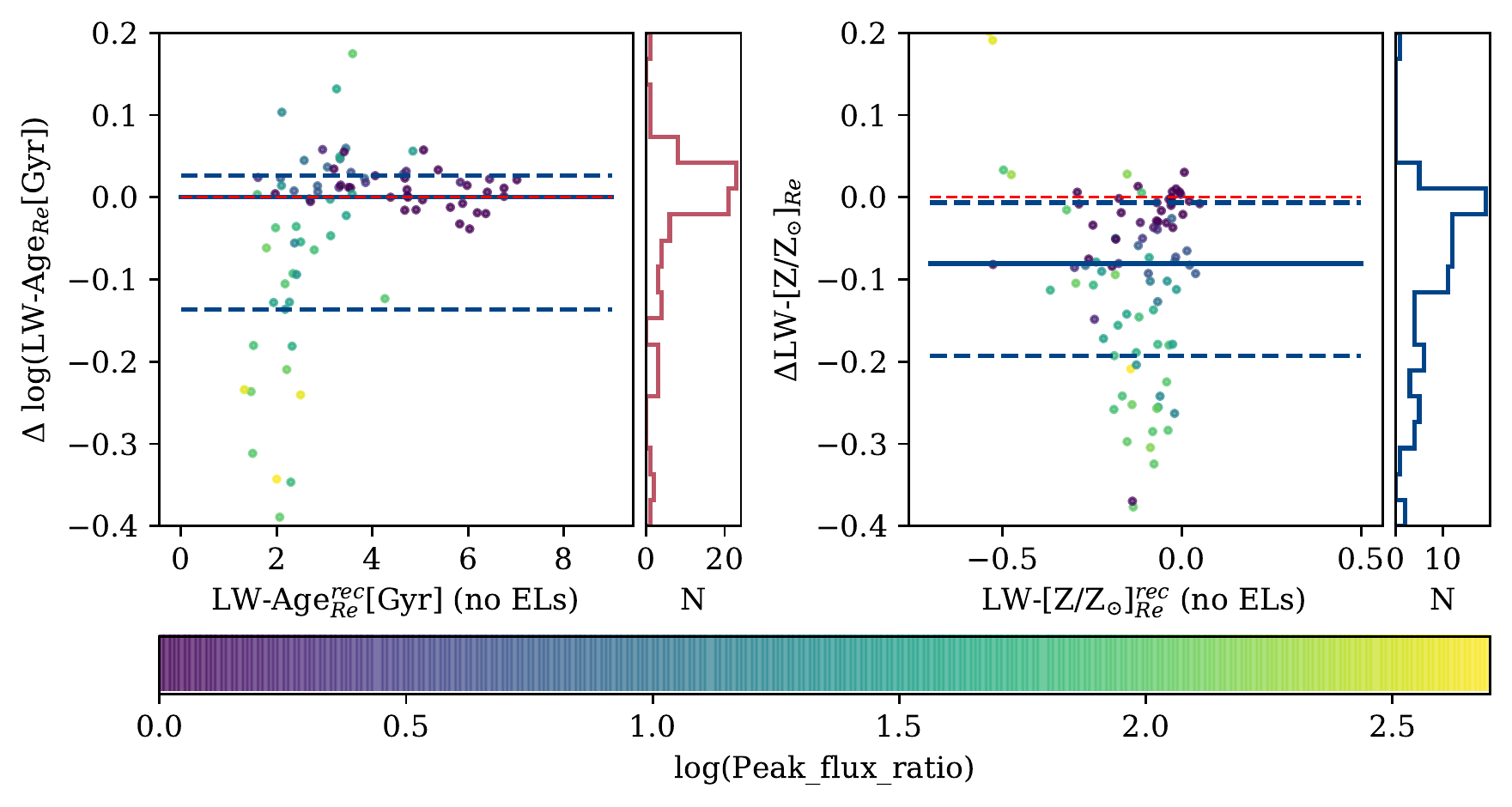}
    \caption{Difference retrieving ages and metallicities (left and right panels, respectively) with \textsc{pyPipe3D} introduced by including emission lines in the spectra. Colour-coding is determined by the peak flux ratio between the spectra with and without the emission lines. The median of the differences is shown with a solid blue line, while the $20^{\rm th}$ and $80^{\rm th}$ percentiles are shown with dashed blue lines.}
    \label{fig:EL-bias}
\end{figure}

Similar to the analysis in Sect.\,\ref{Subsec:pipe3d-bias}, we plot the difference between the mean ages and metallicities derived from the spectra within $1R_{\rm e}$ including ELs and those that do not include ELs for the same galaxies. As we are interested in the effect of ELs in the spectra and some of our $100$ galaxies may have only weak ELs or no ELs, we identify the peak flux ratio between the spectra with and without the ELs within $R_{\rm e}$ for each galaxy. While galaxies with weak ELs seem unaffected (scatter limited to $\pm0.5$\,dex), galaxies with strong ELs tend to lead to younger and more metal-poor estimations than when ELs are not included (Fig.\,\ref{fig:EL-bias}). While this effect is systematic in metallicity, age estimations are more robust but with a larger scatter. The most extreme cases have a deviation of $|\Delta \log ({\rm Age})|\sim 0.4$. At $2\,{\rm Gyr}$, where the deviations are maximum and more frequent, the ages can be estimated to be $\sim1.2\,{\rm Gyr}$ younger. The correlation of this effect with age is evident, since galaxies with young stellar populations will have young massive stars capable of ionizing the surrounding gas, thus producing emission lines.

While this test shows how emission lines can impact in the SSP fitting, a more detailed study on the performance of \textsc{pyPipe3D} is beyond the scope of this work.

\section{Metallicity excess in stellar particles}\label{Ap:met-excess}

The metallicities of TNG50 stellar particles may have values above the upper limit of the \texttt{MaStar\_sLOG} template. These are not properly represented by the template, resulting in a underestimation of the average metallicity of the galaxy. We analyze how frequently the simulated galaxies have particles with metallicities above the threshold given by $Z_{\rm thres}={\rm max}(Z_{\rm SSP})+\Delta Z$, where ${\rm max}(Z_{\rm SSP})=0.4$ is the maximum metallicity ($Z$) in the template and $\Delta Z=0.05$ is half the difference between the highest metallicity in the template and the one immediately before. We consider that particles with metallicities above $Z_{\rm thres}$ are underestimated when constructing the mock datacube.

\begin{figure}
    \centering
    \includegraphics[width=0.5\hsize]{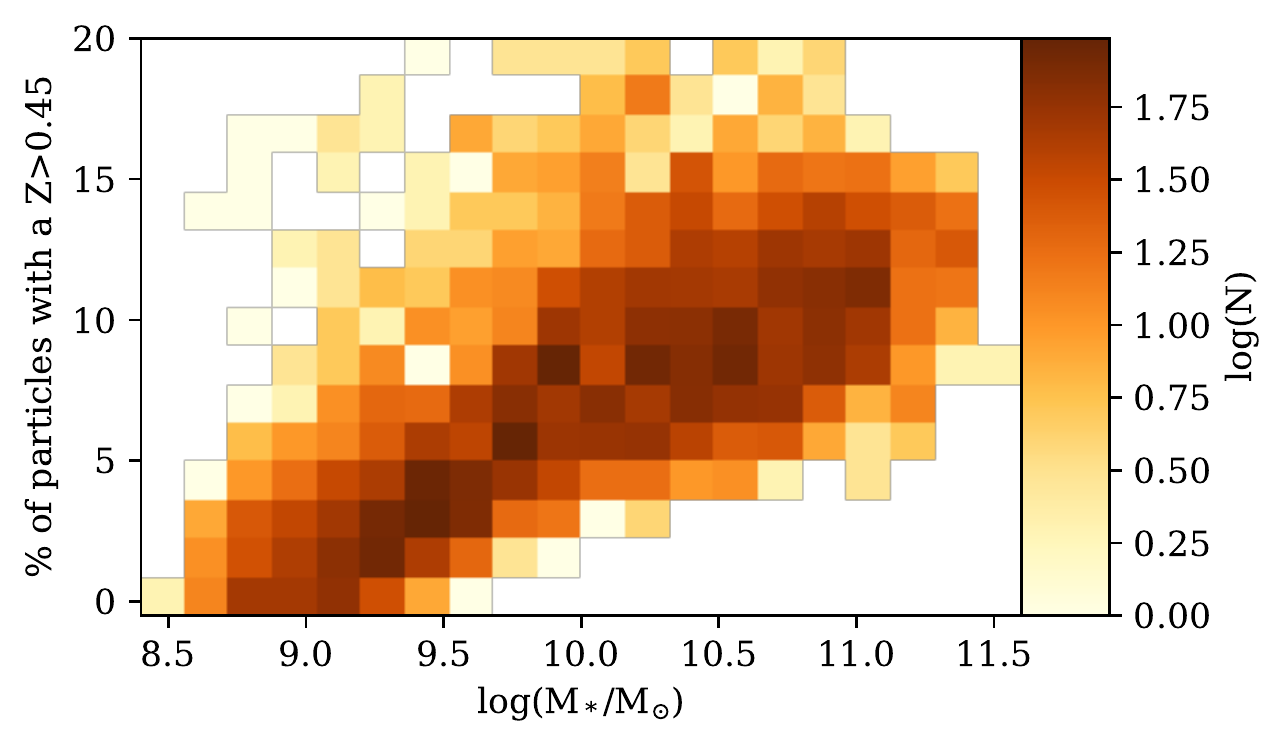}
    \caption{Fraction of stellar particles per galaxy in the MaNGIA sample where the metallicity exceeds $Z_{\rm thres}$(see text) and its dependence with stellar mass. $N$ is the number density of galaxies per bin.}
    \label{fig:met_excess}
\end{figure}

Fig.\,\ref{fig:met_excess} shows the fraction of stellar particles per galaxy in the MaNGIA sample where the metallicity exceeds $Z_{\rm thres}$ and its dependence with stellar mass. For each galaxy, we have considered all the particles identified with the \texttt{subfind} algorithm, which may not coincide with those in the field of view of the mock datacube. There is a clear dependence between stellar mass and high metallicity.

\end{appendix}

\twocolumn
\bibliographystyle{aa}
\bibliography{biblio.bib}

\end{document}